\documentclass[%
reprint,
amsmath,amssymb,
aps,
longbibliography,
nofootinbib
]{revtex4-1}

\usepackage{hyperref}
\usepackage{graphicx}
\usepackage{color}

\usepackage{bm}
\usepackage{amsmath}
\usepackage{natbib}
\usepackage{textgreek}
\usepackage{mathtools}
\usepackage{filecontents}
\usepackage{footmisc}

\DeclarePairedDelimiter{\norm}{\lVert}{\rVert}

\begin{document}
	
	\preprint{APS/123-QED}
	
	\title{The specificity and robustness of long-distance connections \\ in weighted, interareal connectomes}
	
	\author{Richard F. Betzel$^1$}
	\email{rbetzel @ seas.upenn.edu}
	\author{Danielle S. Bassett$^{1,2}$}
	\email{dsb @ seas.upenn.edu}
	\affiliation{
		$^1$Department of Bioengineering, University of Pennsylvania, Philadelphia, PA, 19104 USA}
	\affiliation{
		$^2$Department of Electrical and Systems Engineering, University of Pennsylvania, Philadelphia, PA, 19104 USA}
	
	\date{\today}
	
	\begin{abstract}
	Brain areas' functional repertoires are shaped by their incoming and outgoing structural connections. In empirically measured networks, most connections are short, reflecting spatial and energetic constraints. Nonetheless, a small number of connections span long distances, consistent with the notion that the functionality of these connections must outweigh their cost. While the precise function of these long-distance connections is not known, the leading hypothesis is that they act to reduce the topological distance between brain areas and facilitate efficient interareal communication. However, this hypothesis implies a non-specificity of long-distance connections that we contend is unlikely. Instead, we propose that long-distance connections serve to diversify brain areas' inputs and outputs, thereby promoting complex dynamics. Through analysis of five interareal network datasets, we show that long-distance connections play only minor roles in reducing average interareal topological distance. In contrast, areas' long-distance and short-range neighbors exhibit marked differences in their connectivity profiles, suggesting that long-distance connections enhance dissimilarity between regional inputs and outputs. Next, we show that -- in isolation -- areas' long-distance connectivity profiles exhibit non-random levels of similarity, suggesting that the communication pathways formed by long connections exhibit redundancies that may serve to promote robustness. Finally, we use a linearization of Wilson-Cowan dynamics to simulate the covariance structure of neural activity and show that in the absence of long-distance connections, a common measure of functional diversity decreases. Collectively, our findings suggest that long-distance connections are necessary for supporting diverse and complex brain dynamics.
	\end{abstract}

	\maketitle
	
	
	\section*{Introduction}
	
	The functional repertoire available to any given brain region is shaped by its structural connections \cite{zeki1988functional, sporns2000theoretical, hilgetag2000anatomical, stephan2000computational, passingham2002anatomical}. The complete set of all areas and all connections comprises a \emph{connectome} \cite{sporns2005human}, which can be represented as a network and analyzed using tools from network science \cite{rubinov2010complex}. The network-based approach for studying neural systems is central to the growing field of network neuroscience \cite{bassett2017network}, which seeks to uncover the architectural principles by which the brain is organized, and to both generate and test hypotheses of how the brain's structure supports its function.
	
	Among the most salient organizational features of brain networks is their cost-efficient spatial embedding. Across scales and species, neural elements are arranged such that the brain's \emph{wiring cost} -- the total length of its connections -- is small \cite{buzsaki2004interneuron, chen2006wiring, rivera2011wiring, kaiser2006nonoptimal, horvat2016spatial, rubinov2016constraints}. Low wiring cost helps curtail the material and metabolic expense of forming, using, and maintaining connections, and is thought to offer evolutionary advantages across species \cite{raichle2006brain, van2016comparative}. But despite favoring short-range, low-cost connections, brain networks also exhibit a small proportion of long, costly connections, potentially conferring additional functionality. The opposing drives to reduce wiring cost and promote functionally adaptive structural topology may allow nervous systems to maintain function with a low energy budget \cite{laughlin2003communication, bullmore2012economy, chen2013trade}.
	
	The precise function of long-distance connections is a matter of debate. According to the most widely accepted account, long-distance connections act as bridges to reduce the topological distance between brain areas, thereby facilitating rapid and efficient interareal communication \cite{sporns2004small, van2011rich, bassett2016small}. Though widespread, this account is unsatisfactory for two reasons. First, with advances in imaging and reconstruction techniques, it has become clear that connection weights decay monotonically with interareal Euclidean distance \cite{ercsey2013predictive, rubinov2015wiring, shih2015connectomics, roberts2016contribution}. As a result, the most efficacious communication pathways -- networks' shortest weighted paths -- involve predominantly strong, short-range connections \cite{mivsic2015cooperative, avena2017path}. Second, reductions in topological distance can occur in a non-specific manner: any long-distance connection that reduces topological distance is as good as any other, irrespective of its origin or termination point. Yet, recent empirical evidence in contrast indicates that the brain's long-distance architecture is conserved across and replicable within individuals, suggesting a high level of connectional specificity \cite{heiervang2006between, bassett2011conserved, cammoun2012mapping, markov2012weighted}.
	
	If long-distance connections are not simply random topological shortcuts, what are they? Here, we address this question through analysis of five weighted interareal network datasets representing mouse, \emph{Drosophila}, macaque, and human (high- and low-resolution) connectomes. First, we characterize the spatio-structural architecture of brain networks, demonstrating remarkable consistency across species. Drawing upon decades of research in theoretical neuroanatomy, we demonstrate that clustering brain areas based on their connections' spatial statistics recapitulates aspects of the brain's intrinsic functional network organization, suggesting a spatio-structural basis for brain function. Next, we show that the brain's most efficacious communication pathways -- its shortest weighted paths -- are dominated by short-range connections, undermining the hypothesis that it is the brain's long-distance connections that reduce its average topological distance. Instead, we hypothesize that long-distance connections introduce diverse inputs and outputs to specific brain areas, in the process promoting dynamical complexity. In support of this hypothesis, we demonstrate the dissimilarity of the connectivity profiles of brain areas' long-distance and short-range neighbors, and that long-distance connectivity profiles form clusters, suggesting that the brain's long-distance architecture is both specific and also robust. Finally, using dynamical simulations, we show a reduction in the diversity of functional profiles when long-distance connections are removed, whereas the opposite is true when we remove short-range connections. These findings help clarify the functional roles of the brain's long-distance network architecture and inform future studies investigating network structure and function.

	\begin{figure*}[t]
		\begin{center}
			\centerline{\includegraphics[width=1\textwidth]{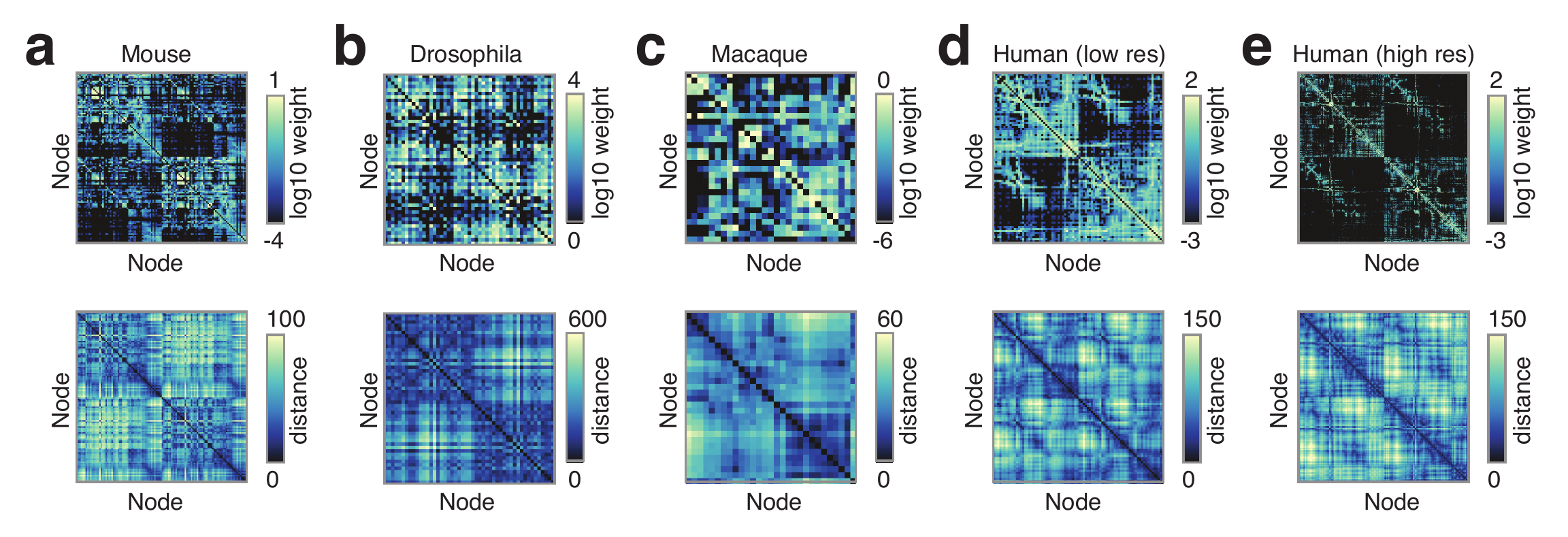}}
			\caption{\textbf{Network matrices.} (\emph{top row}): Matrix representations of connectomes reconstructed from macaque, \emph{Drosophila}, mouse, and human imaging data. (\emph{bottom row}): Euclidean distance matrices for the same species.} \label{fig1}
		\end{center}
	\end{figure*}

	\section*{Results}
	
	\begin{figure*}[t]
		\begin{center}
			\centerline{\includegraphics[width=1\textwidth]{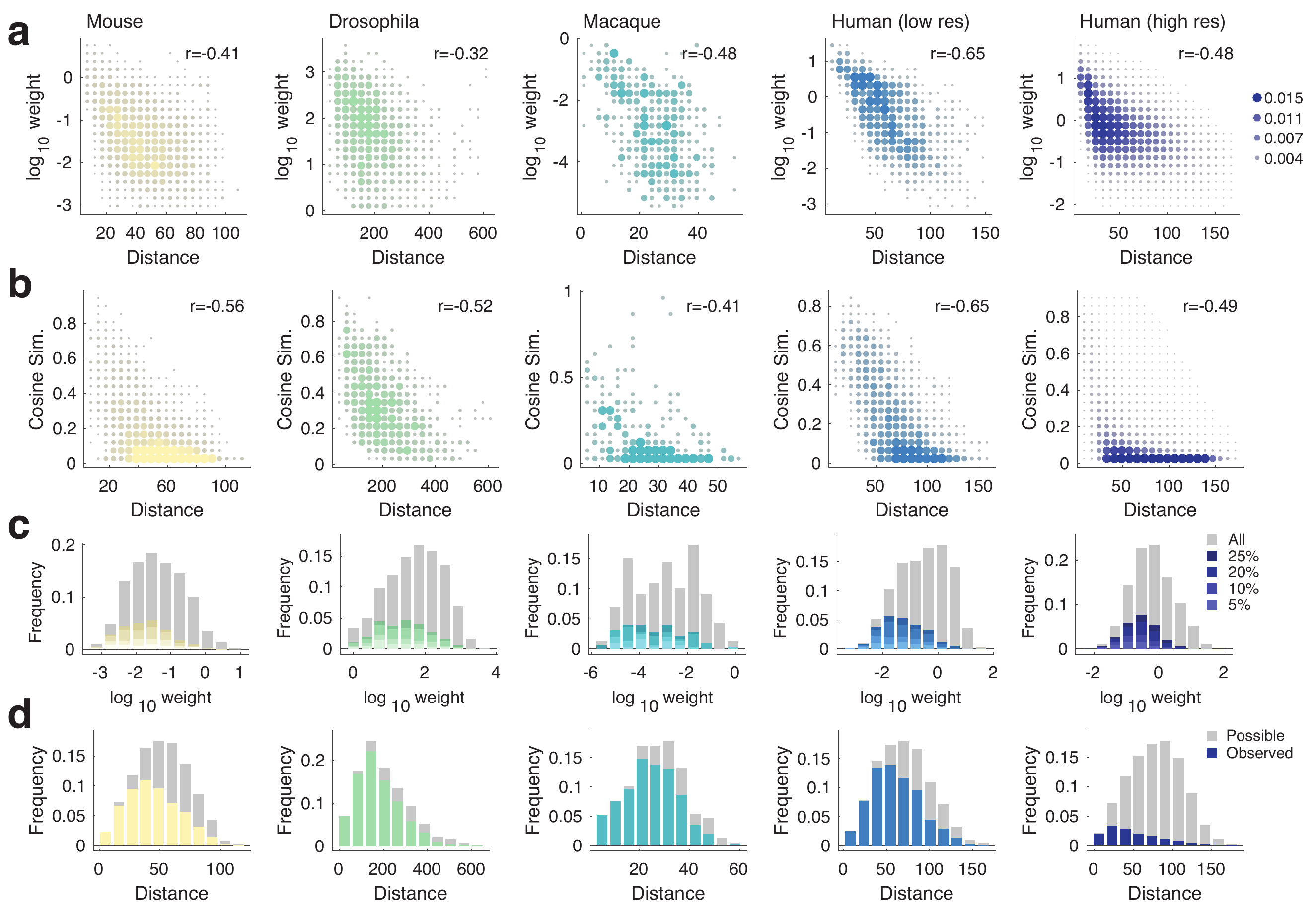}}
			\caption{\textbf{Network distance dependence.} (\emph{a}) Edge weight \emph{versus} distance. (\emph{b}) Cosine similarity \emph{versus} distance. (\emph{c}). Frequency of edge weights across all connections (gray) and long-distance connections (color; top 25\% longest connections). (\emph{d}). Frequency of connection lengths (gray) and lengths of existing connections (color).} \label{fig3}
		\end{center}
	\end{figure*}
	
	Past studies of unweighted brain networks have reported that long-distance connections act to reduce the topological distance between brain areas, supporting efficient communication among distant brain areas \cite{sporns2004small, kaiser2006nonoptimal}. We argue, however, that with new empirical data about connections' weights, this functional interpretation must be revisited. In its place, we propose an alternative set of functional roles for long-distance connections, building on the intuition that brain areas inherit functionality from their patterns of incoming and outgoing connections \cite{zeki1988functional, passingham2002anatomical, betzel2017diversity}. We claim that the primary function of long-distance connections is to deliver unique inputs to brain areas and serve as novel targets for brain areas' outputs, thereby enhancing the functional diversity of those regions. We further hypothesize that long-distance connections are not ``one-offs,'' but instead are insulated and reinforced by other long-distance connections. This architecture naturally leads to increased robustness. In support of these hypotheses, we performed a number of computational experiments involving five interareal connectivity datasets representing four different organisms: mouse, \emph{Drosophila}, macaque, and human (low-resolution and high-resolution). We describe the results of these experiments in the following sections (Fig.~\ref{fig1}). Details of network reconstruction are provided \textbf{Materials and Methods} section.

	\begin{figure*}[t]
		\begin{center}
			\centerline{\includegraphics[width=1\textwidth]{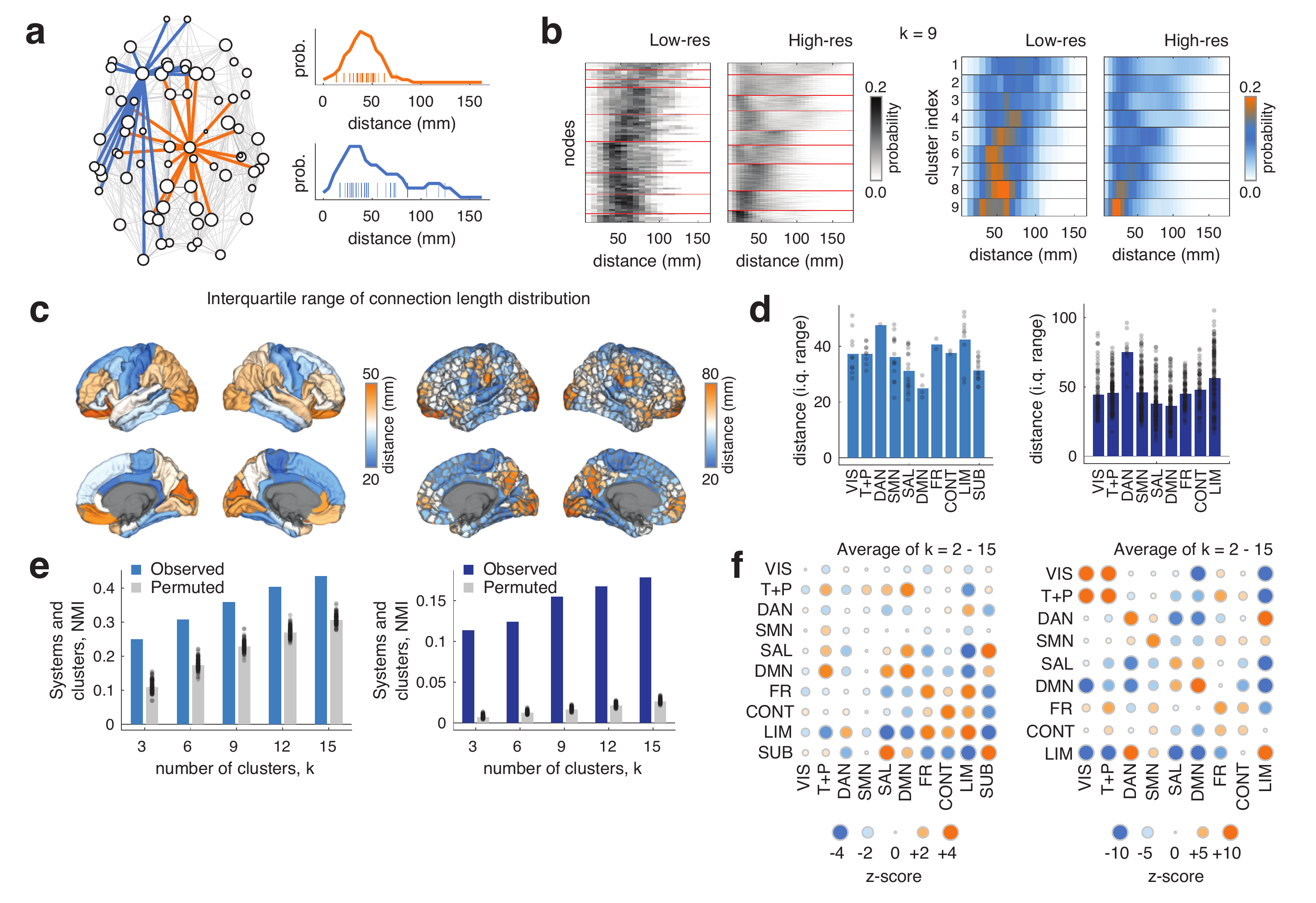}}
			\caption{\textbf{Regional connection length profiles relate to functional specificity in the human network dataset.} (\emph{a}): Schematic illustrating connection length profiles for two example nodes. The orange node makes mostly short- and mid-range connections, while the blue node exhibits some long-distance ($>$100 mm) connections. (\emph{b}) Connection length distributions for a low- and a high-resolution human network dataset. Nodes are ordered according to the cluster to which they were assigned using a $k$-means algorithm. (\emph{c}) Interquartile range of brain areas' connection length distributions. (\emph{d}) Interquartile range plotted for each functional system: Visual (VIS), temporal + precuneus (T+P), dorsal attention (DAN), somatomotor (SMN), salience (SAL), default mode (DMN), frontal (FR), control (CONT), limbic (LIM), and sub-cortex (SUB; applies only to the low-resolution dataset). (\emph{e}) Similarity of $k$-means partitions with functional system labels. (\emph{f}) Standardized ($z$-score) overlap of clusters with functional systems. Circle size indicates the absolute value of the $z$-score and color indicates the sign of the $z$-score. Large orange circles indicate that areas within pairs of systems were more likely to be co-clustered based on their connection length distributions than expected by chance (permutation model).} \label{fig4}
		\end{center}
	\end{figure*}

	\vspace{0.3cm}
	
	\noindent \textbf{Distance shapes weighted network architecture}. We sought to demonstrate that many network-level properties of weighted interareal networks depend upon distance. We focused on four properties in particular: (\emph{i}) connection weight, (\emph{ii}) cosine similarity of connectivity profiles, (\emph{iii}) frequency of long-distance connections among connections of different lengths, and (\emph{iv}) connection probability as a function of distance.
	
	First, we computed the Pearson correlation of the logarithm of connections' weights with Euclidean distances (Fig.~\ref{fig3}a). We observed strong negative correlations across all species and scales (maximum $p < 10^{-15}$; FDR-corrected). Second, we computed the pairwise cosine similarity among all areas' connectivity profiles and computed its correlation with Euclidean distance (Fig.~\ref{fig3}b). As before, we found that cosine similarity was negatively correlated with Euclidean distance (maximum $p < 10^{-15}$; FDR-corrected). Third, we estimated the distributions of the logarithm of connection's weights, and separately labeled the top 5,\%, 10\%, 20\%, and 25\% of all connections by length. We then identified within each histogram bin the contributions made by long connections relative to the contributions made by connections of any length (Fig.~\ref{fig3}c). We observed that the weakest connections were most often associated with the shortest connections, while the strongest connections almost always excluded short connections. Finally, we computed the distribution of all possible interareal Euclidean distances; that is, the elements of the Euclidean distance matrix. Within each histogram bin, we identified which of those possible connections existed and compared to those that did not (Fig.~\ref{fig3}d). We found that when possible, short-range connections were almost always observed, whereas many of the possible long-distance connections were not observed.
	
	Collectively, these results highlight the powerful role that interareal Euclidean distance plays in shaping the structural organization of weighted interareal brain networks. The consistency of these relationships across five datasets is remarkable considering the range of acquisition and reconstruction techniques, and the gross differences in binary network density (the fraction of existing connections irrespective of weight out of all possible connections, $\rho = \frac{\text{\# observed}}{\text{\# possible}}$).
	
	\vspace{0.3cm}
	
	\noindent \textbf{Similarity of connection length distributions shapes areal function}. The functionality of brain areas depends on the configuration and weights of their incoming and outgoing connections \cite{zeki1988functional, passingham2002anatomical}. However, these network properties are correlated with and shaped by distance. It follows, then, that the spatial embedding of a brain area and the lengths of its connections indirectly shape its function. Here, we tested this hypothesis directly by comparing areas' connection length distributions with their functional system assignments.
	
	First, we computed each area's connection length distribution (Fig.~\ref{fig4}a,b). Distributions showed rich topography and varied widely across the cortex. Some were focused and sharply-peaked, while others were broad and included connections of different lengths. To quantify an area's diversity of connection lengths, we computed the interquartile range of its connection length distribution (Fig.~\ref{fig4}c). Broadly, we found that interquartile range varied across functional systems, with dorsal attention and limbic networks exhibiting the greatest levels of diversity in both low- and high-resolution network datasets (Fig.~\ref{fig4}d).
	
	Next, we clustered areas using a k-means algorithm by treating their connection length distributions as features. We implemented the algorithm with 100 random restarts, and we varied the number of clusters from $k = 2$ to $k = 15$. We compared areas' cluster and functional system assignments using a \emph{normalized mutual information} (NMI), where larger NMI values indicate greater overall similarity of clusters. In comparing the observed NMI values with those obtained by randomly permuting areas' cluster assignments, we observed that NMI was consistently greater than expected by chance (1000 permutations; $p < 10^{-3}$, FDR-corrected) (Fig.~\ref{fig4}e). Finally, to determine which functional systems were responsible for driving this similarity, we computed the co-cluster density between every pair of systems. This density represents the fraction of times that pairs of brain areas within those systems were co-assigned to the same cluster by the $k$-means algorithm (Fig.~\ref{fig4}f). In general, we find that pairs of areas within systems are more likely to be co-clustered than expected by chance. With the exception of the visual system in the low-resolution dataset, all $z$-score mean co-cluster densities were greater than zero and with the exceptions of low-resolution visual, dorsal attention, somatomotor, and salience networks, all $z$-scores were statistically significant ($p < 0.05$, FDR-corrected).
	
	Collectively, these results demonstrate that functional specialization of a brain area can be associated with the diversity of its connections and their lengths. This observation suggests that areas with dissimilar connection length distributions tend to have different functional roles within the network, allowing us to ascribe functional significance to connections and their lengths.
	
	\vspace{0.3cm}
	
		\begin{figure*}[t]
		\begin{center}
			\centerline{\includegraphics[width=1\textwidth]{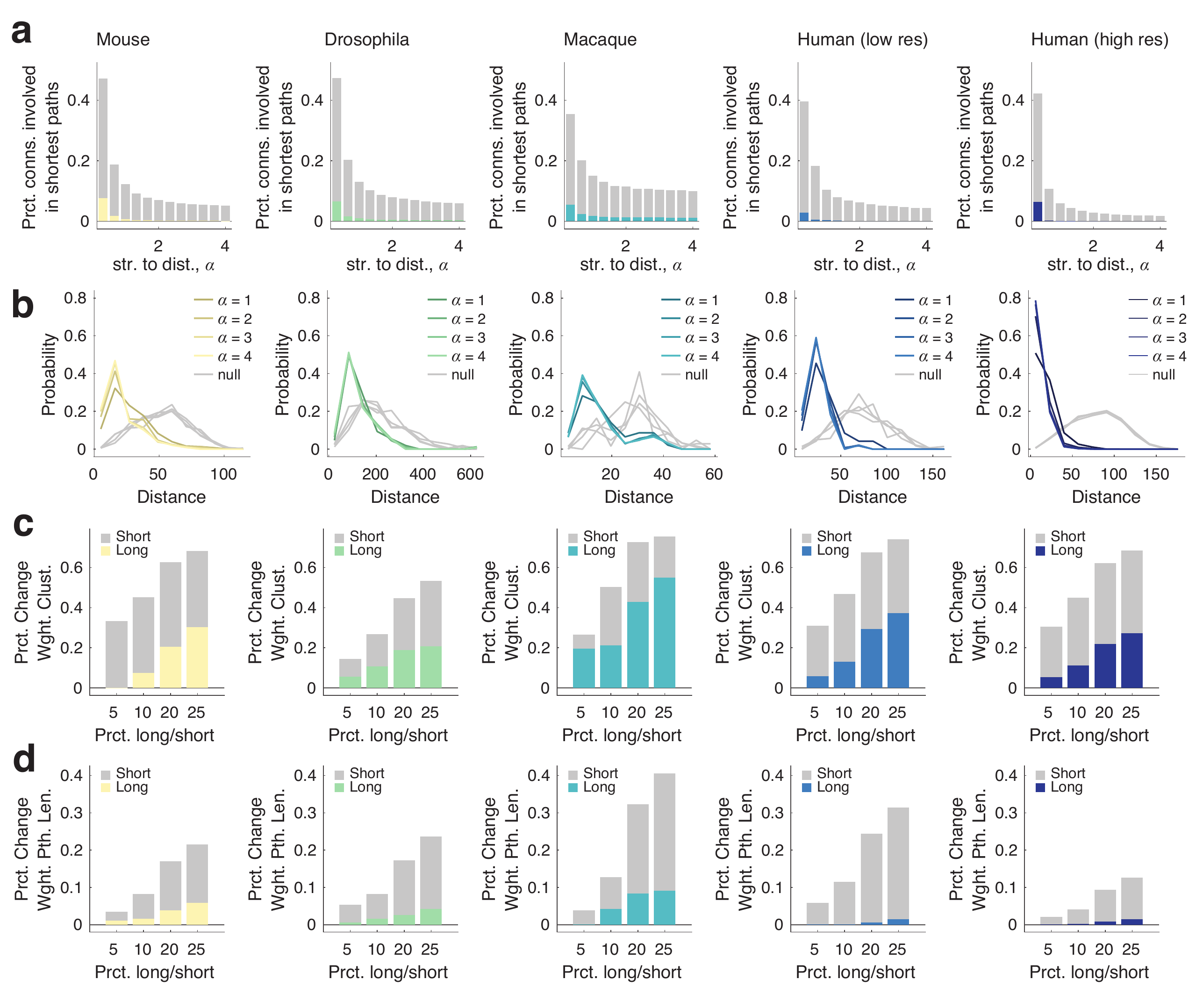}}
			\caption{\textbf{Shortest-path usage in weighted interareal networks.} (\emph{a}) The fraction of total connections used in shortest paths. The total fraction is shown in gray; the long-distance fraction (top 25\% length) is shown in color. The \emph{x}-axis represents the edge strength-to-distance parameter, $\alpha$. (\emph{b}) Edge length distributions of connections participating in shortest paths. Gray curves show the mean distribution under a permutation-based null model; colored curves show $\alpha = 1,2,3,4$. (\emph{c}) Percent change in weighted clustering coefficient as a result of removing different fractions of long and short connections. (\emph{d}) Percent change in weighted characteristic path length as a result of removing different fractions of long and short connections. Note: bar plots in panel \emph{d} are shown with $\alpha = \frac{1}{3}$. At larger values of $\alpha$ long-distance connections play no role in shortest path structure and removing them leads to no change in the weighted characteristic path length.} \label{fig5}
		\end{center}
	\end{figure*}
	
	\noindent \textbf{Long connections contribute little to shortest, weighted paths}. According to current literature, long-distance connections play important roles in networks' shortest-path structure and reduce the number of processing steps between brain areas \cite{sporns2004small}. This observation, however, was made using binary networks where edges carry no weight \cite{muldoon2016small}. It is less clear what role long-distance connections play when connections are weighted and when those weights span multiple orders of magnitude. We hypothesized that, due to the disparity between the strongest and weakest connection weights and their dependence on distance, the network's shortest weighted paths would be dominated by short-range connections.
	
	To test this hypothesis, we computed a parameterized version of edge betweenness centrality, $BC_{ij}(\alpha)$. The $\alpha$ parameter, which we varied in increments of $1/3$ over the range $[0,4]$, controlled the decay rate of the necessary mapping of connection weights to length. For each network and for each value of $\alpha$, we obtained an estimate of $BC_{ij}(\alpha)$, the fraction of shortest paths that contained the connection $\{i,j\}$. Then, we calculated the fraction of long-distance connections (top 25\% by length) involved in at least one shortest path. We observed that long-distance connections played a minor role when $\alpha$ was small, but an increasingly small role as $\alpha \rightarrow 4$ (Fig.~\ref{fig5}a). We investigated this behavior further by computing the distribution of connection distances for all pairs of brain areas that participated in at least one shortest path (Fig.~\ref{fig5}b). We found that across all datasets, the observed distributions were skewed towards short-range connections, while a null model in which connection topology was preserved but brain area locations were randomly permuted exhibited a broader distribution that involved many more long-distance connections ($p < 10^{-5}$; FDR-corrected). These results demonstrate that long-distance connections play relatively minor roles in the shortest-path structure of weighted interareal brain networks. Because a network's shortest paths are interpreted as routes along which brain areas communicate with one another (see, however, \cite{goni2014resting, avena2017path} for alternative perspectives), these findings suggest that routing information along high-weight pathways composed of short-range connections is more efficient than using weak, long paths for interareal communication.
	
	\begin{figure*}[t]
		\begin{center}
			\centerline{\includegraphics[width=1\textwidth]{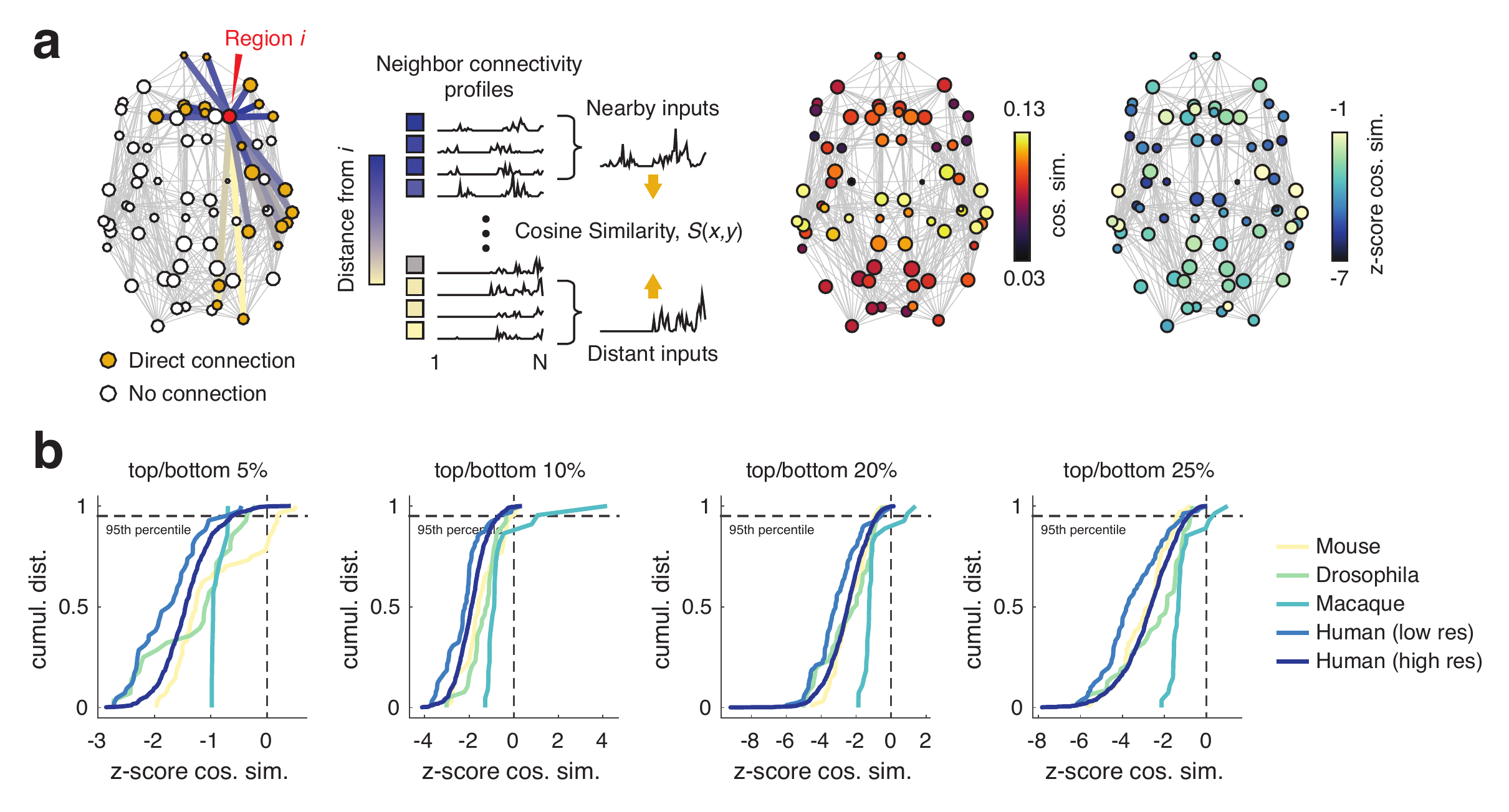}}
			\caption{\textbf{Similarity of long- and short-range connectivity profiles.} (\emph{a}) Schematic of processing pipeline for assessing the similarity of long- and short-range inputs. The network depicted is the human, low-resolution network, thresholded to a binary density of $\rho = 0.25$ and with subcortical areas removed for visualization purposes only. For the empirical analysis, we performed no thresholding and retained all areas in all computations. All connections incident upon area $i$ are identified and their lengths are tabulated. Node $i$'s neighbors are then classified as either nearby or distant. Note: the same distance threshold was applied uniformly to all brain areas. Separately, the connectivity profiles of nearby \emph{versus} distant neighbors are summed. The summed profiles, which represent possible inputs to $i$ from its neighbors, are compared to one another using the cosine similarity measure. This process results in a single similarity score for each area (node). We compare these scores against a null distribution obtained by randomly re-classifying neighbors as nearby \emph{versus} distant. (\emph{b}) Cumulative distributions of area-level $z$-scores for each network. The different panels represent variation of the threshold for classifying neighbors as nearby \emph{versus} distant. From \emph{left} to \emph{right}, nearby (distant) neighbors were those connected by the top (bottom) 5\%, 10\%, 20\%, and 25\% of connection lengths.} \label{fig6}
		\end{center}
	\end{figure*}

	Next, and for completeness, we demonstrated that removing short connections has a much greater impact on statistics related to small worldness than removing the same fraction of long connections. Specifically, we calculated the percent change in mean weighted path length and mean weighted clustering coefficient (Fig.~\ref{fig5}c,d)\footnote{Note: The percent change in weighted clustering coefficient and path length are shown as absolute values.}. We systematically varied our definition of short \emph{versus} long connections, focusing on the shortest \emph{versus} longest 5\%, 10\%, 20\%, and 25\% of connections according to the Euclidean distances. For all datasets, we observed that the effect of removing strong, short connections was consistently greater than that of removing long connections. These results paint a picture in which nervous system function and communication is dominated by strong, low-cost structural connections. These findings are inconsistent with the view of nervous system function in which most communication pathways are funnelled through a small proportion of long-distance connections.
	
	\vspace{0.3cm}
	
	\begin{figure*}[t]
		\begin{center}
			\centerline{\includegraphics[width=1\textwidth]{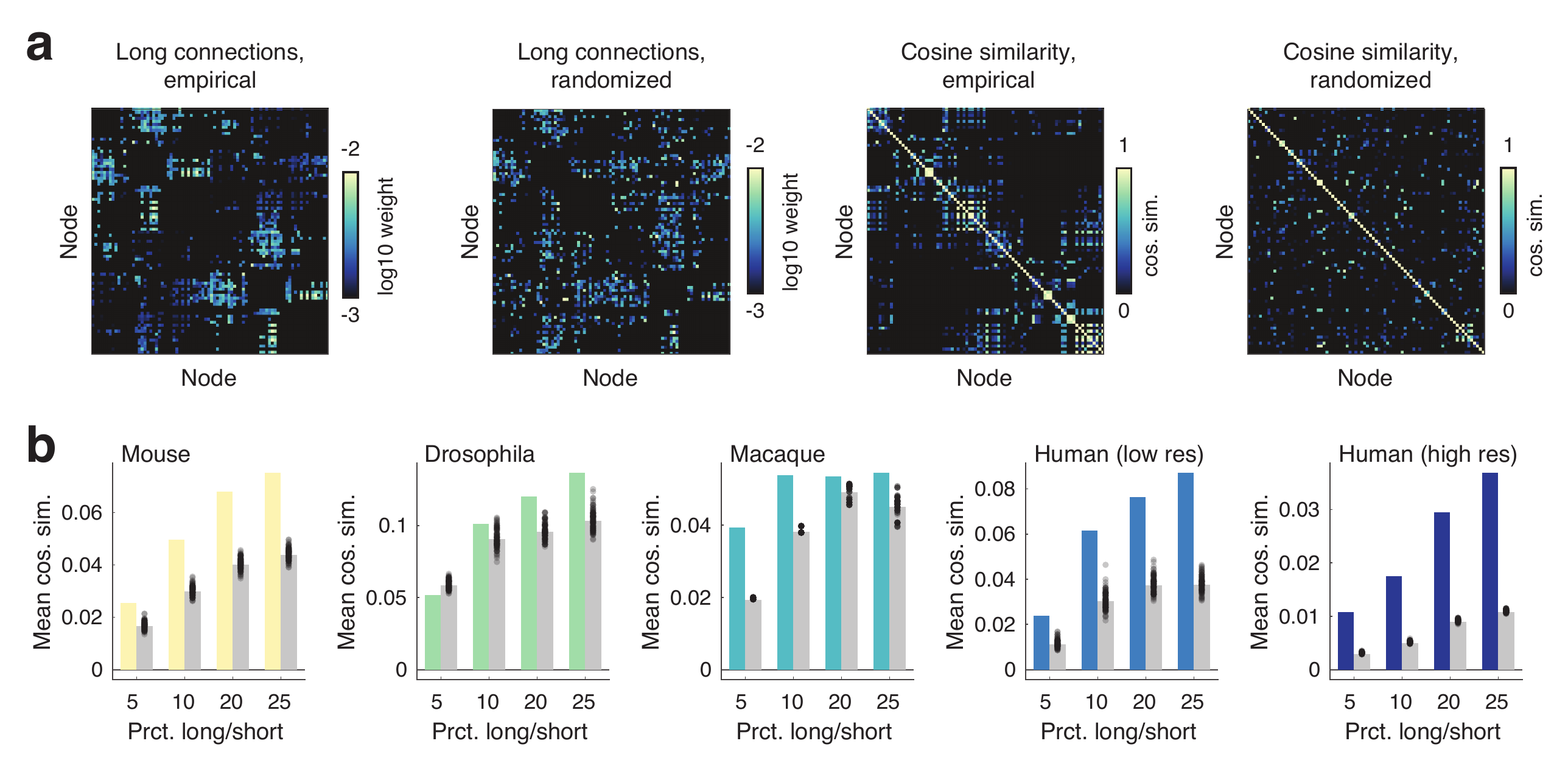}}
			\caption{\textbf{Redundancy of long-range connectivity.} (\emph{a}) Example empirical and randomized networks thresholded to retain the 25\% longest connections (\emph{leftmost} panels). We then compute the pairwise cosine similarity between areas' long-range connectivity profiles (\emph{rightmost} panels). (\emph{b}) Pairwise similarity measures are averaged for the empirical and randomized networks. We repeat this process using four different definitions of ``long-range'': 5\%, 10\%, 20\%, and 25\% longest connections as defined by Euclidean distance between regional center of mass.} \label{fig7}
		\end{center}
	\end{figure*}
	
	\noindent \textbf{Long and short connections deliver dissimilar inputs and outputs}. If long-distance connections play little or no role in the shortest weighted path structure of brain networks, what do they contribute? To understand the functional roles of long-distance connections, we compared them against short-range connections in terms of the character of connectivity profiles. An area's connectivity profile specifies the other areas it can influence and be influenced by, thereby shaping its functional properties.
	
	We considered the neighbors of brain area $i$ and examined the average connectivity profile of those nearest and those most distant. We hypothesized that, compared to short-range connections, long-distance connections would deliver unique inputs to and novel targets for an area, $i$, and as a result their respective connectivity profiles should be dissimilar. We quantified this dissimilarity with cosine similarity, and compared the value observed in the empirical data to a null distribution generated by keeping the network topology fixed but permuting areas' spatial locations (Fig.~\ref{fig6}a). We computed the similarity of each areas' long-distance and short-range neighbors' connectivity profiles, while varying the definition of long \emph{versus} short connections (top/bottom 5\%, 10\%, 20\%, and 25\%). We observed that the distribution of standardized similarity scores was consistently negative, indicating that long-distance and short-range connections are more dissimilar from one another than expected by chance (Fig.~\ref{fig6}b). More quantitatively, we observed that the cumulative distribution of standardized similarity scores reached 95\% before a positive value was encountered (with the exception of the macaque and in one case the mouse dataset). These results confirm that the pattern of incoming and outgoing connections to brain areas are dissimilar when compared to one another on the basis of their lengths. This observations suggests that a wealth of long-distance connections may enhance an area's functional repertoire, by providing unique inputs as well as novel targets for output.
	
	 \begin{figure*}[t]
	 	\begin{center}
	 		\centerline{\includegraphics[width=1\textwidth]{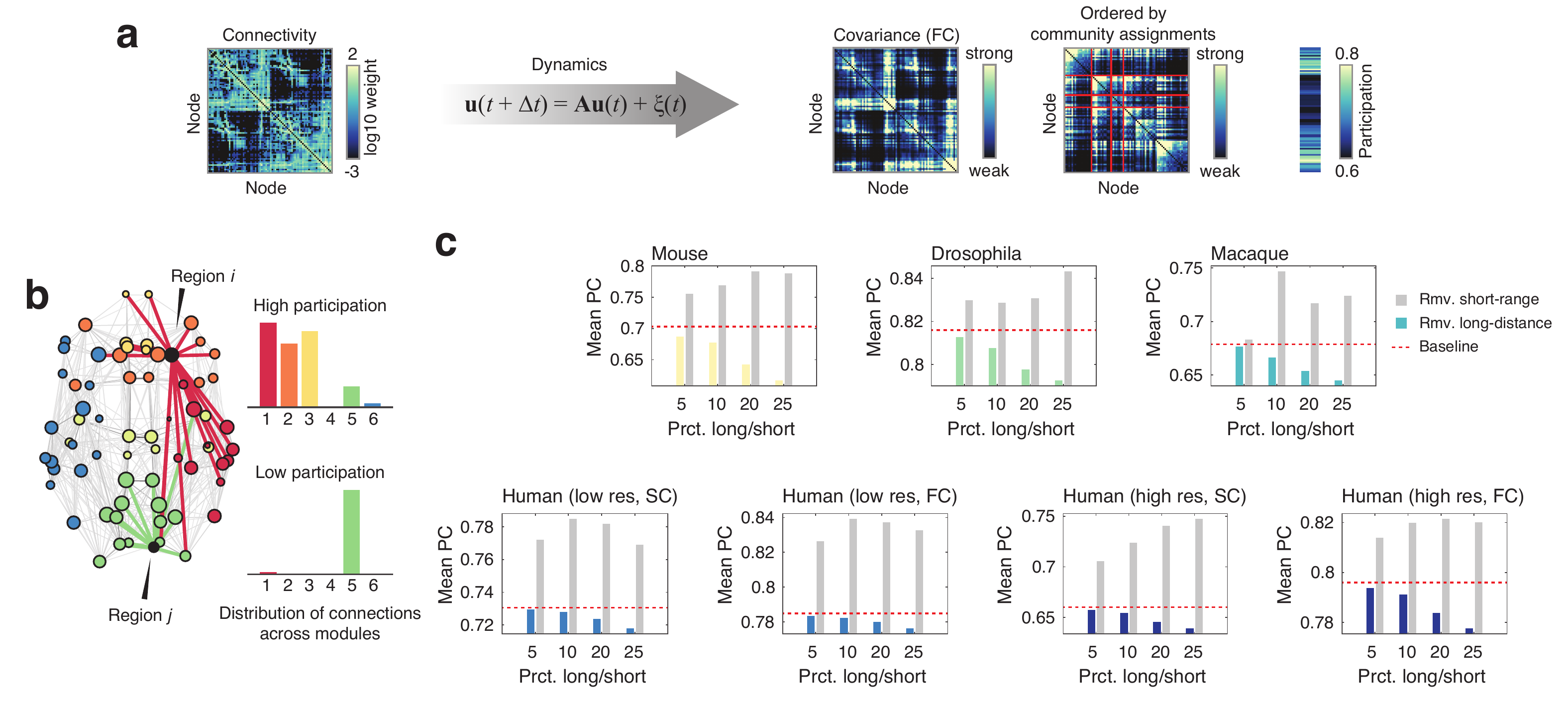}}
	 		\caption{\textbf{Linearized dynamics and participation coefficient.} (\emph{a}) Schematic illustrating analysis pipeline. Structural connectivity matrix is used to constrain a linearization of Wilson-Cowan dynamics, which results in an estimated covariance (FC) matrix. We compute network modules ahead of time and, based on those modules and on the simulated covariance structure, we compute brain areas' functional participation coefficients. (\emph{b}) Schematic illustrating the concept of the participation coefficient (PC). An area with high participation coefficient (red connections) forms connections to many different modules, while the connections of an area with low participation coefficient (green connections) are largely restricted to a single module. An area's participation coefficient can be interpreted as a measure of its connectional diversity. When computed using a covariance matrix, an area's participation coefficient measures its functional diversity. (\emph{c}) Change in mean participation coefficient after removing different percentages of a network's longest and shortest connections. The mean participation of the intact network is depicted as a red dashed line. Note that removing long-distance connections consistently reduces the mean participation coefficient (colored bars), indicating a decrease in functional diversity. Removing short-range connections (gray bars) has the opposite effect. Note that for the low- and high-resolution human datasets, we computed the participation coefficient with respect to structural and functional modules.} \label{fig8}
	 	\end{center}
	 \end{figure*}

	 \vspace{0.3cm}
	
	 \noindent \textbf{Long-distance connections exhibit degeneracies}. Short-range and long-distance connections lead to neighbors with dissimilar connectivity profiles. However, there are many scenarios in which this could occur. For instance, long-distance connections could be dissimilar not only from short-range connections, but also from other long-distance connections. That is, each long-distance connection links area $i$ to another area whose connectivity profile is not similar to the connectivity profile of any of $i$'s other long-distance neighbors. In this scenario, each long-distance connection provides a truly unique set of inputs and output targets with respect to one another. This uniqueness, however, could belie a weakness; damage to a single connection would result in the absence of inputs or outputs, to or from $i$. Another possibility is that long-distance connections are reinforced in some way with built-in degeneracy. That is, from $i$'s perspective, its long-distance connections lead to areas with similar connectivity profiles, so that if one or a small number of connections were damaged, then $i$ would still receive similar inputs and deliver similar outputs.
	
	 To test whether this was the case, we imposed distance thresholds on connectivity matrices so that connections below a certain length were excluded (Fig.~\ref{fig7}a). We computed the pairwise similarity of each area's long-distance connectivity profile, and we then computed the mean similarity over all pairs. Larger mean similarity scores are indicators of increased levels of degeneracy. In parallel, we compared these empirical scores against those obtained from a randomized null model in which a network's degree sequence and edge weight distribution were exactly preserved, and in which a network's connection length distribution and length-weight relationship were preserved approximately (code available at: \url{https://www.richardfbetzel.com/code/}). We repeated this process as we varied the definition of ``long-distance'' (the top 5\%, 10\%, 20\%, and 25\% according to connection lengths). We observed that the mean similarity of long-distance connectivity profiles consistently exceeded that of the null model (non-parametric test, $p < 0.01$; FDR-corrected). The lone exception was observed in the \emph{Drosophila} dataset when using the most exclusive definition of long-distance (top 5\%). Otherwise, this result was observed across all network datasets.

     These findings indicate that the long-distance architecture of brain networks exhibits non-random, correlations. This observation is inconsistent with recently-proposed stochastic models of brain network formation in which connections of all lengths are generated by the same wiring rules \cite{ercsey2013predictive, song2014spatial}. Instead, areas' long-distance connections are organized in such a way that many brain areas exhibit highly similar patterns of incoming and outgoing connections.
	
	 \vspace{0.3cm}
	
	 \noindent \textbf{Long-distance connections lead to diverse patterns of functional coupling}. The results presented in the previous sections described the structural architecture of brain networks. Though we interpreted these results in terms of brain function and information transmission, the link between structure and function is complicated, and recent findings have cast doubt on the role of shortest paths in network communication \cite{goni2014resting, mivsic2015cooperative}. In this section, we ground these intuitions in a dynamical model of neural activity and demonstrate that long-distance connections support the functional diversity of brain areas.
	
	 Specifically, we use a linearization of the Wilson-Cowan population model \cite{galan2008network} and compute from this model the covariance matrix of brain areas' temporal activity. The rows of this matrix represent brain areas' \emph{functional fingerprints} -- their pattern of functional coupling to the $N - 1$ other brain areas (Fig.~\ref{fig8}a). We define an area's functional diversity to be equal to its participation coefficient (PC) computed given its corresponding covariance matrix and given modules estimated from the structural connectivity network using modularity maximization (Fig.~\ref{fig8}b). The functional diversity of the whole brain is defined as the average PC over all brain areas. We then repeated this procedure, removing the same number of short-range and long-distance connections, and computed the resultant change in average PC. Across all network datasets, we found that removing long-distance connections always resulted in decreased average PC, while removing short-range connections resulted in increased average PC. This finding suggests that reductions in the number of long-distance connections in the network will tend to decrease the overall functional diversity of the brain, resulting in a more specialized and less integrated network as a whole.

	\section*{Discussion}
	
	Brain areas' functional repertoires are shaped by their incoming and outgoing structural connections. Most connections are short-range and incur minimal cost to the network in terms of material and energy. Nonetheless, a small proportion of connections span long distances, leading to the hypothesis that the functionality conferred by these connections must outweigh their cost. Their precise function, however, is a matter of debate. The current hypothesis is that long-distance connections reduce the average topological distance between brain areas, facilitating interareal communication.
	
	Here, we challenge this hypothesis on the grounds that it is not necessarily true for weighted brain networks and that it implies a non-specificity of long-distance connections. We propose, instead, that long-distance connections serve to diversify brain areas' inputs and outputs, and to promote complex network dynamics. To test this hypothesis, we analyze five weighted interareal brain network datasets. As expected, we find that brain areas' long-distance and short-range neighbors exhibit marked differences in terms of their connectivity profiles, suggesting that long-distance connections contribute to the specificity of regions' connectivity profiles and serve as sources of dissimilar inputs and outputs for brain areas. Next, we show that in isolation areas' long-distance connectivity profiles exhibit non-random levels of similarity, suggesting that the communication pathways formed by long connections exhibit redundancies, which may help promote robustness. Finally, we simulate the covariance structure of neural activity using a linearization of Wilson-Cowan dynamics. We show that in the absence of long-distance connections, one measure of functional diversity -- the mean participation coefficient -- decreases, indicating that from a functional perspective, long-distance connections are necessary for supporting diverse and complex brain dynamics.
	
	\vspace{0.3cm}
	
	\noindent \textbf{Interpreting the functional roles of long-distance connections}. One of the open challenges of theoretical neuroanatomy is understanding how brain function is shaped by structure \cite{kotter2001neuroscience}. Long-distance connections, because they are prevalent despite high material and metabolic cost, are believed to play critical functional roles. The earliest network analyses argued that long-distance connections acted as integrative structures that reduced topological distance between brain areas, which contributed to efficient interareal communication \cite{sporns2004small}. This perspective was largely based on analogy between brain and other socio-technical networks whose functions are better-understood \cite{watts1998collective}.
	
	In the decade and a half since these early analyses, however, this narrative has largely remained unrefined (though there have been some recent advances. See, for example, \textcite{bassett2016small}). In general, long-distance connections are almost universally regarded as key integrative structures. Recently, however, the functional roles of long-distance connections have been revisited. In \textcite{markov2013role}, the authors demonstrated that in the macaque brain, the similarity of areas' connectivity profiles decreased with distance, suggesting that long-distance connections contributed to an area's specificity. More recently and using a different macaque network dataset, \textcite{chen2017features} suggested that long-distance connections, those unanticipated by a model that penalized the formation of costly connections, form a dense cluster, and may also perform some segregative functions.
	
	Our work builds on these and other recent papers investigating the functional consequences of the brain's spatial embedding and the roles of costly, long-distance connections \cite{samu2014influence, betzel2017modular, chen2013trade, ercsey2013predictive, horvat2016spatial, roberts2016contribution, song2014spatial, betzel2016generative}. Our findings show that, across species and scales, interareal communication along shortest paths is dominated by strong short-range connections, with long-distance connections contributing minimally. Instead, we find that brain areas' neighbors \emph{via} long-distance connections have massively different connectivity profiles than their short-range neighbors. This finding is a clear demonstration that, from a structural perspective, long-distance connections allow brain areas to interact through novel configurations of inputs and outputs. The consistency of this observation across species is also suggestive of an evolutionarily-conserved mechanism of interareal communication.
	
	\vspace{0.3cm}
	
	\noindent \textbf{The specificity of long-distance connections}. In our work, we also showed that brain areas' profiles of long-distance connections were similar to one another, a feature not accounted for by degree sequence, distance and weight distributions, and weight-distance relationships. This observation suggests that the organization of long-distance connections (and possibly connectomes as a whole) is shaped by an underlying latent structure that is a function of brain areas' spatial locations but also some set of unknown factors, including cytoarchitectonic and transcription profile similarity \cite{hilgetag2016primate, beul2015predictive, richiardi2015correlated}, higher order topological organization \cite{henriksen2016simple, betzel2016generative}, or temporal staging in which connections are formed during developmental windows when brain areas are proximal to one another \cite{kaiser2017mechanisms}. The similarity of areas' long-distance connectivity profiles also suggests a sense of connectional specificity that may not be explained by recent papers that proposed stochastic and globally-enforced wiring principles \cite{ercsey2013predictive, song2014spatial, horvat2016spatial}. While these and other models can provide insight into brain-wide organizational principles \cite{vertes2012simple, betzel2016generative, henriksen2016simple, betzel2017generative}, they oftentimes lack the ability to accurately predict area-level statistics \cite{chen2017features}.
	
	\vspace{0.3cm}
	
	\noindent \textbf{Interpreting connection weights}. The process of enumerating a network's shortest paths does not incorporate information about spatial relationships. The observation that shortest paths are dominated by short-range connections is a direct result of the inverse relationship between connection weight and distance, as well as the assumption that connections' weights represent the communication efficacy between connected brain areas. This interpretation, however, exposes a shortcoming common in most empirical analyses of connectome data. Namely, that the weights of connections are estimated from structural data alone, and while we ascribe functional significance to their values, may have no true bearing on brain function or communication (e.g., the number or fraction of reconstructed streamlines or projections does not imply that they are consistently used for signaling). Are we justified, then, in our functional interpretation of connection weights?
	
	At the scale of brain areas, communication is determined in part by axonal diameter and myelination status, which place limits on nerve conduction velocity \cite{hursh1939conduction, ritchie1982relation}. Interestingly, a number of studies have reported roughly lognormal-distributed fiber diameters \cite{buzsaki2014log}, which is in broad agreement with the connection weight distributions reported here. Moreover, due to volumetric \cite{rivera2011wiring} and cost constraints \cite{bullmore2012economy}, the probability of axonal projections spanning long distances is small \cite{laughlin2003communication, markov2010weight}. So while the distributions of connection weights are both consistent across datasets and agree with previously reported results in which other variables relevant to communication were also measured, it still remains unclear whether the weights can be directly interpreted in terms of interareal communication. Ultimately, addressing this question remains an empirical challenge, but would be of tremendous practical value as it would inform network modeling efforts, both of brain structural networks \cite{betzel2017generative} but also of its function \cite{deco2008dynamic, leon2013virtual, chaudhuri2015large}.

	\vspace{0.3cm}
	
	\noindent \textbf{Limitations}. This study has a number of methodological limitations. First, we assume that we can make meaningful claims about the functional properties of nervous systems and brain areas by studying their structural networks alone. The validity of this assumption is built upon decades of empirical observations and recent simulation studies demonstrating that the covariance structure of temporally evolving neural activity can be predicted from properties of the structural matrix \cite{honey2009predicting, goni2014resting, abdelnour2014network, becker2015accurately}. Our claims, however, are more nuanced than simply stating that function and structure are related to one another. We claim, specifically, that the shortest path structure of brain networks is of functional importance, a view that has been challenged of late \cite{goni2014resting, avena2017path}. While the precise role of shortest paths is, indeed, unclear, disruptions to shortest paths have been associated with disease, suggesting that an important, if poorly understood, functional role, and thereby motivating their study, here \cite{crossley2014hubs, korgaonkar2014abnormal}.
	
	Second, the networks studied here were composites built from many single-subject observations. It remains unclear to what extent these networks are representative of the average individual. Moreover, it is important to note that in the absence of connection weights, long-distance and short-range connections are effectively weighted the same. In this extreme case, it is likely that networks' shortest paths \emph{will}, in fact, include many long-distance connections. We can begin to see this when we tune the parameter $\alpha$ closer to zero. In any case, the ``correct'' edge-weighting scheme is unknown. Lastly, diffusion imaging and tractography, exhibit known biases that make it challenging to detect long-distance cortico-cortical tracts \cite{reveley2015superficial, sotiropoulos2017building}. While future methodological advances may prove helpful \cite{pestilli2014evaluation}, our results are bolstered by the inclusion of a high-resolution multiband diffusion imaging scan marking 257 diffusion directions.

	\section*{Conclusion}
	We present evidence that long-distance connections are not merely topological shortcuts. Instead, they introduce diversity among brain areas' neighbors, which we show in human data can be related to brain function. We also confirm using simulations of brain dynamics that in the absence of long-distance connections, brain networks exhibit a decrease in their functional diversity. Lastly, long-distance connections exhibit degeneracies, so that many different areas have similar patterns of long-distance connectivity. We speculate that this degeneracy confers robustness to the system. Our findings contribute to a growing body of literature aimed at refining our understanding of how brain structure shapes its function.

	\section*{Materials and Methods}
	We analyzed mouse, \emph{Drosophila}, macaque, and human weighted, interareal network datasets. Each dataset was distinct in terms of imaging modality, reconstruction technique, and connection weighting scheme. This variability in processing strategy was unintentional, though we exploit this feature in order to demonstrate the universality of our findings and their robustness to acquisition and processing schemes. In this section, we describe the methods used to reconstruct and analyze the networks.
	
	\subsection*{Network datasets}
	\noindent \textbf{Mouse}. The mouse connectivity matrix was reconstructed based on freely available tract-tracing data from the Allen Institute Mouse Brain Connectivity Atlas (\url{http://connectivity.brain-map.org}; see \cite{oh2014mesoscale} for more details of tract-tracing experiments). Anterograde recombinant adeno-associated viral tracer was injected into target areas in the right hemisphere of mouse brains, which was extracted three weeks post-injection at which time viral tracer projection patterns were reconstructed. Reconstructions were then smoothed and aligned to a common coordinate space of the Allen Reference Atlas.
	
	Network nodes were defined according to a custom parcellation based on the Allen Developing Mouse Brain Atlas \cite{rubinov2015wiring}. This parcellation contains 65 areas in each hemisphere, 9 of which were removed because they were not involved in any tract-tracing experiment. The resulting weighted and directed network contained $N = 112$ areas of interest linked by edges corresponding to interareal axonal projections and weighted as normalized connection densities: the number of connections from unit volume of a source area to unit volume of a target area (Fig.~\ref{fig1}a).
	
	\vspace{0.3cm}
	
	\noindent \textbf{\emph{Drosophila}}. The connectivity matrix for \emph{Drosophila} was reconstructed from the FlyCircuit 1.1 database (\url{http://www.flycircuit.tw}), a repository of images of 12,995 projections neurons in the female \emph{Drosophila} brain \cite{chiang2011three}. Neurons were labeled with green fluorescent protein (GFP) using genetic mosaic analysis with a repressible cell marker. GFP-labeled neurons were delineated from whole-brain three-dimensional images and co-registered to a female template brain using a rigid linear transform. Individual neurons were partitioned into $N = 49$ local processes units (LPUs; network nodes) with distinct morphological and functional characteristics. LPUs were defined so as to contain their own population of local interneurons whose fibers were limited to that LPU. The result was a weighted and directed network comprised of projections among LPUs. This network has been analyzed elsewhere \cite{shih2015connectomics, betzel2017diversity, worrell2017optimized} (Fig.~\ref{fig1}b).
	
	\vspace{0.3cm}
	
	\noindent \textbf{Macaque}. The macaque connectivity matrix was based on retrograde tract-tracing experiments and originally reported in \cite{markov2012weighted}. Injections of fluorescent tracers were made in 28 macaque monkeys. Reconstructed projections were localized with respect to a parcellation comprised of $N = 91$ cortical areas based on histological and atlas-based landmarks. For each tract-tracing experiment, the number of labeled neurons in each of the 91 areas was counted. This number was then expressed relative to the number of labeled neurons minus the number of neurons intrinsic to the injection site. The result is a $[29 \times 91]$ matrix of connection weights from each injection site to the rest of the brain. We focused on the $[29 \times 29]$ weighted and directed connectivity matrix \cite{markov2013role} (Fig.~\ref{fig1}c).
	
	\vspace{0.3cm}
	
	\noindent \textbf{Human structural networks}. Human brain networks were reconstructed from diffusion weighted magnetic MRI using deterministic tractography algorithms. The networks we analyzed were group-representative composites of subject-level networks (30 subjects). This network construction process entailed acquiring diffusion spectrum and T1-weighted anatomical images for each individual. DSI scans sampled 257 directions using a Q5 half-shell acquisition with a maximum $b$-value of 5000, an isotropic voxel size of 2.4 mm, and an axial acquisition with repetition time $TR = 5$ seconds, echo time $TE = 138$ ms, 52 slices, and field of view of $[231,231,125]$ mm. All procedures were approved by the Institutional Review Board of the University of Pennsylvania.
	
	DSI data were reconstructed in DSI Studio (\url{www.dsi-studio.labsolver.org})), using $q$-space diffeomorphic reconstruction (QSDR) \cite{yeh2011estimation}. QSDR reconstructs diffusion-weighted images in native space, computes the quantitative anisotropy (QA) of each voxel, warps the image to a template QA volume in Montreal Neurological Institute (MNI) space using the statistical parametric mapping nonlinear registration algorithm, and reconstructs spin-density functions with mean diffusion distance of 1.25 mm with three fiber orientations per voxel. Fiber tracking was performed using a modified FACT algorithm with an angular cutoff of $55^\circ$, step size of 1.0 mm, minimum length of 10 mm, spin density function smoothing of 0.00, maximum length of 400 mm, and a QA threshold determined by DWI signal in the colony-stimulating factor. For each individual, the algorithm terminated when 1,000,000 streamlines were reconstructed.
	
	In parallel, T1 anatomical scans were segmented using FreeSurfer and parcellated using the Connectome Mapping Toolkit (\url{http://www.connectomics.org}) according to low- and high-resolution atlases ($N_{low} = 82$ and $N_{high} = 1000$) \cite{cammoun2012mapping}. The low-resolution atlas comprised 68 cortical areas and 14 subcortical structures. The high-resolution atlas comprised 1000 cortical areas, representing subdivisions of cortical areas delineated in the low-resolution atlas, and 14 subcortical structures. Note that the upsampling procedure applied to the cortical areas was not applied to the subcortical structures. As a result, the volumes and surface areas of subcortical structures in the high-resolution atlas were many times greater than that of the high-resolution cortical areas. Because large morphometric disparities can induce unwanted biases in network analysis, we elected to exclude sub-cortical structures from our analysis of networks constructed using the high-resolution atlas. Each parcellation was registered to the B0 volume of subjects' DSI data, and a B0-to-MNI voxel mapping was used to map area labels from native space to MNI coordinates. Streamlines were aggregated by the areas in which their starting and terminal endpoints were located. The connection weight between any pair of areas was defined as their streamline count normalized by the geometric means of their volumes.
	
	Group-representative matrices were generated using a distance-dependent, consistency-based thresholding procedure. This procedure was applied separately to inter- and intra-hemispheric connections. The resulting networks had a binary density equal to the average across subjects, approximately the same distribution of inter- and intra-hemispheric edge lengths, and approximately the same edge weight distribution as every subject (Fig.~\ref{fig1}d,e). This approach has been described elsewhere \cite{mivsic2015cooperative, betzel2017modular} and shown to be superior to distance-agnostic thresholding procedures \cite{roberts2017consistency}.
	
	\subsection*{Network analysis}
	
	Inter-areal networks were represented as weighted connectivity matrices, $\mathbf{W} \in \mathbb{R}^{N \times N}$, where the element $w_{ij}$ denoted the strength of the connection between brain areas (nodes) $i$ and $j$. We encoded spatial relationships between nodes with Euclidean distance matrices, $\mathbf{E} \in \mathbb{R}^{N \times N}$, where the element $e_{ij}$ denoted the straight-line distance between the physical locations of areas $i$ and $j$.
	
	\vspace{0.3cm}
	
	\noindent \textbf{Shortest weighted paths}. The shortest path between two areas represents the most direct channel by which they can communicate, with shorter paths implying enhanced communication capacity. To calculate the shortest path structure of weighted networks, we first performed a weight-to-length transformation of the network's connections. This procedure was necessary because connection weights measure the affinity of one node to another, while shortest path algorithms seek to minimize a measure of length or cost. One possible transformation is to take the element-wise reciprocal of $\mathbf{W}$, so that the length of the connection between areas $i$ and $j$ is given by $l_{ij} = 1/w_{ij}$. This transformation can be made more general by introducing the parameter, $\alpha > 0$, so that $l_{ij}(\alpha) = w_{ij}^{-\alpha}$. Under this parameterization, $l_{ij}(\alpha)$ decays monotonically as a function of $w_{ij}$ with a rate of decay rate given by $\alpha$. Once weights were transformed into lengths, the network's shortest path structure was computed and stored in the distance matrix, $\mathbf{D} \in \mathbb{R}^{N \times N}$, whose element $d_{st} = l_{si} + l_{ij} + \ldots + l_{kt}$ encoded the length of the weighted shortest path between source area $s$ and target area $t$ \cite{dijkstra1959note}.
	
	\vspace{0.3cm}
	
	\noindent \textbf{Mean weighted path length}. Given a network's shortest path structure, we calculated a number of useful metrics. The simplest was the average length of shortest paths:
	
	\begin{equation}
	\langle L \rangle = \frac{2}{N(N - 1)} \sum_{i,j > i} d_{ij}.
	\end{equation}
	
	\noindent This measure tells us, on average, the cost of using shortest paths for communication.
	
	\vspace{0.3cm}
	
	\noindent \textbf{Edge betweenness centrality}. We also calculated the contributions made to a network's shortest path structure by its connections \cite{brandes2001faster}. Let $\pi_{s \rightarrow t} = \{ \{s,i\}, \{i,j\}, \ldots , \{k,t\} \}$ be the sequence of connections traversed along the shortest path from a source node $s$ to a target node $t$. A connections's betweenness centrality, $BC_{ij}$, measures the fraction of all shortest paths that include the connections $\{i,j\}$; its value can be interpreted as a measure of a connection's importance for communication along a network's shortest paths.
	
	\vspace{0.3cm}
	
	\noindent \textbf{Interareal similarity}. A brain area's functionality is derived from its connectivity profile, i.e. its pattern of incoming and outgoing connections \cite{zeki1988functional, passingham2002anatomical}. The connectivity profile of area $i$ is defined as the vector $\mathbf{w}_i = [w_{i1} , \ldots , w_{iN}]$. Regions with similar connectivity profiles have the capacity to receive and deliver similar input and output signals, and are therefore thought to play roughly equivalent functional roles within the network \cite{hilgetag2002computational, barbas1989architecture, young1993organization, scannell1995analysis}. To measure the functional relatedness of two areas $i$ and $j$, we can calculate the similarity of their connectivity profiles as the cosine of the angle, $\theta_{ij}$, formed by the vectors $\mathbf{w}_i$ and $\mathbf{w}_j$:
	
	\begin{equation}
	S_{ij} = cos(\theta_{ij}) = \frac{\mathbf{w}_i \cdot \mathbf{w}_j}{\norm{\mathbf{w}_i}\norm{\mathbf{w}_j}}.
	\end{equation}
	
	\noindent Note that for directed networks, we define an area's connectivity profile to include both its incoming and outgoing connections. That is, $\mathbf{w}_i = [w_{i1} , \ldots , w_{iN}, w_{1i} , \ldots , w_{Ni}]$.
	
	\vspace{0.3cm}
	
	\noindent \textbf{Short- and long-distance connections}. While a brain area's functionality depends on its own connectivity profile, it also depends on the connectivity profiles of its neighbors. An area's neighbors can have connectivity profiles dissimilar from its own and therefore can contribute unique inputs to that area or offer novel targets for that area's outgoing connections.
	
	We measure the uniqueness of inputs and outputs using cosine similarity. Specifically, we compare the connectivity profiles of area $i$'s neighbors linked by short- and long-distance connections. Let $\Gamma_i = \{j : w_{ij} \ne 0 \}$ be the set of $i$'s neighbors. These neighbors can be subdivided into short- and long-distance subsets: $\Gamma_i^{\text{short}} = \{j : w_{ij} > 0, e_{ij} \le \tau_e^{\text{short}} \}$ and $\Gamma_i^{\text{long}} = \{j : w_{ij} > 0, e_{ij} \ge  \tau_e^{\text{long}} \}$. Here, $\tau_e^{\text{short/long}}$ represent distance cutoffs below or above which we consider a neighbor to be short- \emph{versus} long-distance with respect to $i$'s location.
	
	To demonstrate that $i$'s short- \emph{versus} long-distant neighbors have dissimilar connectivity profiles and therefore unique inputs and outputs, we computed the cosine similarity of their mean connectivity profiles. The mean connectivity profile of $\Gamma_i^{\text{short/long}}$, is defined as:
	
	\begin{equation}
	\mathbf{w}_i^{\text{short/long}} = \sum_{j \in \Gamma_i^{\text{short/long}}} \mathbf{w}_j.
	\end{equation}
	
	\noindent The similarity of $\mathbf{w}_i^{\text{short}}$ and $\mathbf{w}_i^{\text{long}}$, $S_{i^{\text{short}}i^{\text{long}}}$ was compared to a randomized null model, in which the network's topology was kept fixed, but where nodes' locations were randomly permuted. This procedure tests the null hypothesis that the (dis)similarity of connectivity profiles from short- and long-distance neighbors could arise under random spatial embeddings as a result of the network's topology alone.
	
	\vspace{0.3cm}
	
	\noindent \textbf{Redundancy of long-distance connections}. Complex networks are subject to perturbations and their components can degrade over time, processes that compromise network function \cite{albert2000error}. To counter these processes, many systems exhibit structural degeneracy in which a multiplicity of components play the same or similar functional roles \cite{tononi1999measures}. In the event that some of these components are damaged, system function is maintained by the remaining undamaged components. We hypothesized that if the brain's long-distance connections were organized to provide unique and specific inputs and outputs, then these pathways should exhibit structural degeneracy. To test this, we compared whether areas's long-distance connectivity profiles were more similar to one another than expected in a randomized null model in which the network's degree sequence, edge-length distribution, and edge-weight distribution were maintained exactly. This was accomplished by retaining connections and weights whose lengths, $e_{ij} \ge \tau_e^{\text{long}}$. Using this network of long-distance connections only, we computed the cosine similarity for every pair of brain area connectivity profiles and computed the average similarity across all pairs:
	
	\begin{equation}
	\langle S^{\text{long}} \rangle = \frac{2}{N(N - 1)} \sum_{i,j > i} S_{ij}^{\text{long}}.
	\end{equation}
	
	\noindent Larger values of $\langle S^{\text{long}} \rangle$ indicate greater levels of structural degeneracy.
	
	\vspace{0.3cm}
	
	\noindent \textbf{Modularity maximization}. Many complex systems, including brain networks, exhibit rich meso-scale structure such that their nodes can be meaningfully partitioned into clusters \cite{newman2012communities}. Modular organization, in which clusters represent weakly interacting sub-systems called modules or communities, is a well-described phenomenon in both structural and functional brain networks \cite{sporns2016modular}. Here, we use modularity maximization to uncover network modules \cite{newman2004finding}. Modularity maximization seeks to partition nodes into modules such that the intra-modular density of connections maximally exceeds that of a null connectivity model. This is accomplished by heuristically maximizing the modularity quality function:
	
	\begin{equation}
	Q (\gamma) = \sum_{ij} b_{ij} \delta(c_i,c_j).
	\end{equation}
	
	\noindent Here, $b_{ij} = w_{ij} - \gamma \cdot p_{ij}$, where $w_{ij}$ and $p_{ij}$ are the observed and expected weights of the connection between nodes $i$ and $j$. The resolution parameter, $\gamma$, scales the relative importance of $p_{ij}$ and determines the number and size of detected modules. Node $i$'s module assignment is encoded as $c_i \in \{1 , \ldots , K \}$. Here, $\delta(c_i,c_j)$ is the Kronecker delta function and is equal to 1 when $c_i = c_j$ and zero otherwise. Effectively, $Q$ is computed as a summation over node pairs, $\{i,j\}$, that fall within modules and is maximized when these pairs are more strongly connected than anticipated.
	
	\vspace{0.3cm}
	
	\noindent \textbf{Structural network modules}. We applied modularity maximization to each of the five structural network datasets. For structural brain networks, we used a null connectivity model that preserves nodes' binary and weighted degrees but otherwise allowed connections to be formed at random. Under this model $p_{ij} = \frac{k_i^{out}k_j^{in}}{2m}$, where $k_j^{in} = \sum_i w_{ij}$ and $k_i^{in} = \sum_j w_{ij}$. Following recent work, we symmetrized the value of $b_{ij}$ as $b_{ij} = \frac{(w_{ij} - \gamma \cdot p_{ij}) + (w_{ji} - \gamma \cdot p_{ji})}{2}$ \cite{leicht2008community}. Note that for the undirected human networks $k_i^{in} = k_i^{out}$ and so the symmetrization was unnecessary.
	
	We used a generalized version of the Louvain algorithm to maximize $Q(\gamma)$ \cite{jutla2011generalized}, varying $\gamma$ from 0.5 to 4.0 in increments of 0.1. At every value of $\gamma$ we repeated the Louvain algorithm 100 times with random initial conditions. We selected the optimal value of $\gamma$ by computing the pairwise similarity ($z$-score of the Rand index \cite{traud2011comparing}) of partitions and focusing on local maxima of the $\gamma$ \emph{versus} median similarity curve. At local maxima, we generated a representative consensus partition from the partitions produced by the Louvain algorithm (see \cite{betzel2017modular} for more details).
	
	\vspace{0.3cm}
	
	\begin{figure*}[t]
		\begin{center}
			\centerline{\includegraphics[width=0.75\textwidth]{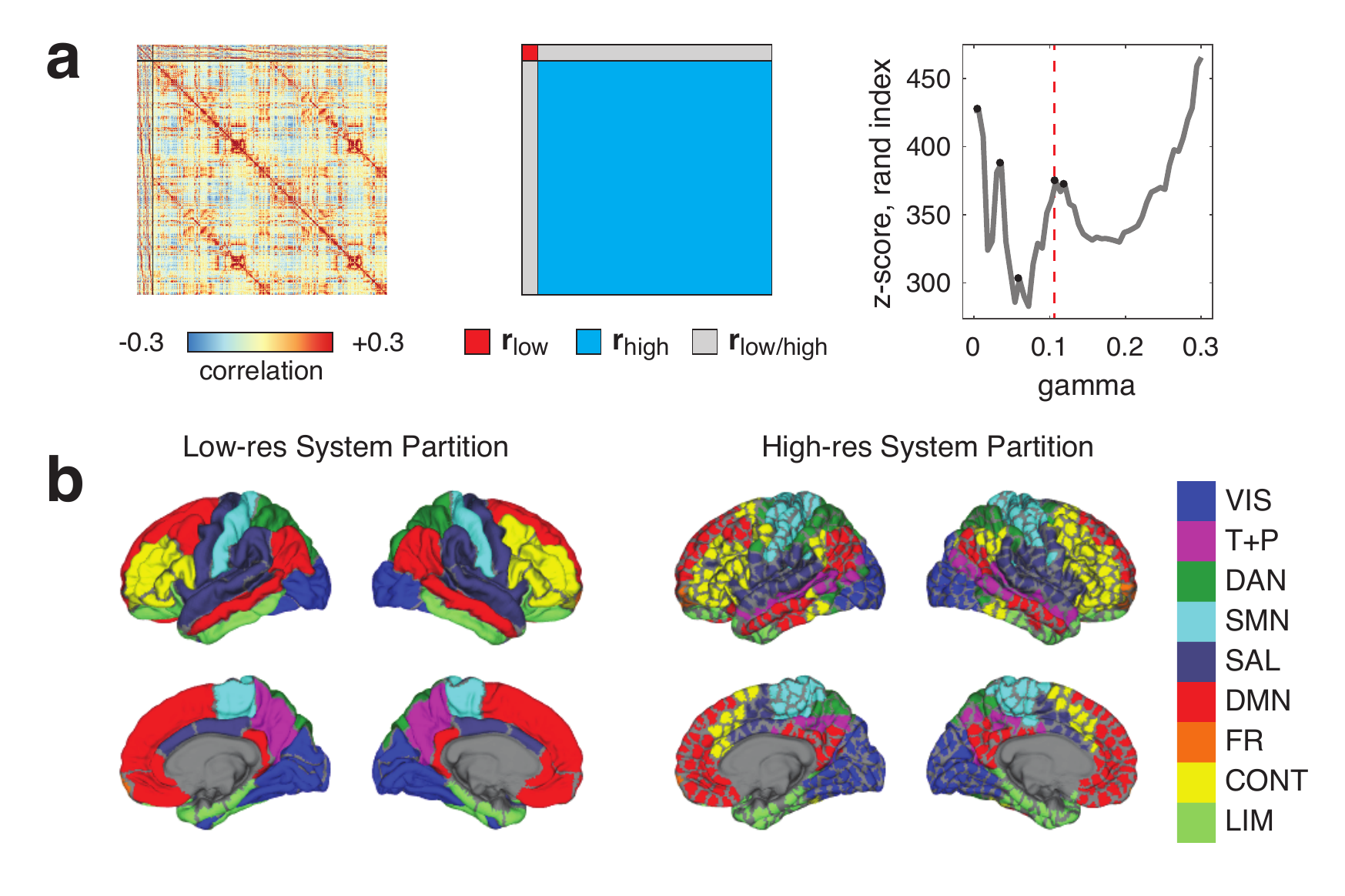}}
			\caption{\textbf{Multi-layer modularity maximization for functional systems.} (\emph{a},\emph{left}) Flattened multi-layer correlation matrix, $\mathbf{r}$. (\emph{a},\emph{middle}) Illustration of multi-layer matrix divided into low- and high-resolution layers and inter-layer coupling. (\emph{a},\emph{right}) Similarity of detected partitions as a function of the resolution parameter, $\gamma$. Local maxima are shown as black circles. The dotted red line depicts the value of $\gamma$ used to construct the system partitions. \emph{b} Spatial topography of low- and high-resolution versions of the system partitions.} \label{fig2}
		\end{center}
	\end{figure*}
	
	\noindent \textbf{Functional network modules}. The focus of this paper was on structural brain networks. However, to facilitate more direct functional interpretation of results obtained from analyses of human networks, we sought estimates of brain areas' functional system assignments. To obtain such estimates, we applied modularity maximization to correlation matrices constructed from independently acquired task-free fMRI BOLD data, the details of which have been described elsewhere \cite{mivsic2015cooperative, betzel2013multi}. Briefly, this dataset comprised 40 subjects that underwent 9-minute resting-state scans, following which BOLD time series were extracted from the same $N_{low} = 82$ and $N_{high} = 1000$ areas as described above and subsequently averaged across subjects. We refer to these group-level correlation matrices as $\mathbf{r}_{low}$ and $\mathbf{r}_{high}$.
	
	Applying modularity maximization to functional brain networks to obtain system labels requires extra care for several reasons. First, establishing a consistent set of system identities across scales is non-trivial because low- and high-resolution functional network datasets are generally constructed and clustered independently from one another. Here, we mitigate this issue using a multi-layer network model and a generalization of modularity maximization to cluster both matrices simultaneously \cite{mucha2010community}. Specifically, we treated $\mathbf{r}_{low}$ and $\mathbf{r}_{high}$ as differently-sized layers inter-linked to one another through the rectangular correlation matrix, $\mathbf{r}_{low/high}$, whose elements encoded the correlation magnitude of activity in low-resolution areas with that of high-resolution areas. The resulting multi-layer network was flattened to have dimensions $\mathbf{r} \in \mathbb{R}^{1068 \times 1068}$, where the first 68 nodes represent cortical areas from the low-resolution functional network and the final 1000 nodes represent cortical areas from the high-resolution network.
	
	Second, the null connectivity model must be compatible with whatever measure was used to define functional connection strength (in this case a Pearson correlation coefficient) \cite{macmahon2013community}. As suggested by \textcite{bazzi2016community}, we defined $p_{ij} = 1$ for all node pairs $\{i,j\}$, so that $b_{ij} = r_{ij} - \gamma$. This expression corresponds to a null model in which BOLD activity of all nodes is uniformly correlated with a magnitude of $\gamma$. As before, the free parameter $\gamma$ determines the number and size of communities. We tested $\gamma \in [0,0.3]$ in increments of 0.006 for a total of 51 possible values. We used the same strategy described earlier to identify $\gamma$ values of interest and to obtain consensus partitions. This analysis resulted in nine modules consistent across low- and high-resolution datasets. Based on spatial topography and visual inspection, we named these modules: visual (VIS), temporal + precuneus (T+P), dorsal attention (DAN), somatomotor (SMN), salience (SAL), default mode (DMN), frontal (FR), control (CONT), and limbic networks (LIM) (Fig.~\ref{fig2}b).
	
	\vspace{0.3cm}
	
	\noindent \textbf{Participation coefficient}. A network's meso-scale organization highlights groups of brain areas thought to perform similar functions. Once those groups were identified, we further characterized the functional roles of individual brain areas based on their structural interactions with modules. One popular measure for doing so is an area's participation coefficient, which measures the extent to which an area's connections are concentrated within a few modules or are distributed more evenly across many modules \cite{guimera2005functional}.
	
	\begin{equation}
	PC_i = 1 - \sum_{s = 1}^K \bigg( \frac{k_{is}}{k_i}\ \bigg)^2.
	\end{equation}
	
	\noindent Here $k_i = \sum_j w_{ij}$ is $i$'s weighted degree and $k_{is} = \sum_{j \in s} w_{ij}$ is the total weight of $i$'s connections to module $s$. Participation coefficients range from 0 to 1, where larger values indicate that connections are evenly spread over modules. An area's participation coefficient can be interpreted as a measure of its diversity of function \cite{bertolero2017diverse}.
	
	\vspace{0.3cm}
	
	\noindent \textbf{Simulated functional connectivity}. The network analyses described above are purely \emph{structural} -- they characterize architectural features of networks whose connections represent physical and material pathways in the brain. While network structure is oftentimes interpreted in terms of a network's function -- e.g., information or signals being routed along a network's shortest paths -- the link can be made stronger by modeling or simulating dynamical processes over the network \cite{woolrich2013biophysical}.

Here, we study a discretized and linearized model of Wilson-Cowan dynamics, as described in \cite{galan2008network} and used in \cite{honey2009predicting}. Let $\mathbf{u}(t) = \{u_1(t) , \ldots , u_N(t) \}$ be the vector of brain areas' states (activity levels) at time $t$. Under this model, states evolve as:
	
	\begin{equation}
	\mathbf{u}(t + \Delta t) = \mathbf{A} \mathbf{u}(t) + \xi(t)
	\end{equation}
	
	\noindent where $\xi(t)$ is uncorrelated Gaussian noise and $\Delta t$ is a single time step. Here, the generalized coupling matrix, $\mathbf{A}$, was defined as:
	
	\begin{equation}
	\mathbf{A} = (1 - \alpha \Delta t) \mathbf{I} + \mathbf{W} \Delta t
	\end{equation}
	
	\noindent where $\alpha$ is a leak variable within each brain area and $\mathbf{I}$ is the identity matrix. As in \textcite{honey2009predicting}, we fixed $\alpha = 2$.
	
	The covariance matrix of areas' states over time can be calculated directly from the spectral properties of the coupling matrix, $\mathbf{A}$, and the covariance of the noise terms, $\xi(t)$. As with empirical covariance matrices, we interpret the covariance matrix computed here as an estimate of functional connectivity. See \textcite{galan2008network} for more details.
	
	\section*{Author Contributions}
	
	This study was designed, carried out, and the manuscript written by RFB and DSB.
	
	\section*{Acknowledgements}
	
	RFB and DSB would like to acknowledge support from the John D. and Catherine T. MacArthur Foundation, the Alfred P. Sloan Foundation, the Army Research Laboratory and the Army Research Office through contract numbers W911NF-10-2-0022 and W911NF-14-1-0679, the National Institute of Health (2-R01-DC-009209-11, 1R01HD086888-01, R01-MH107235, R01-MH107703, R01MH109520, 1R01NS099348 and R21-M MH-106799), the Office of Naval Research, and the National Science Foundation (BCS-1441502, CAREER PHY-1554488, BCS-1631550, and CNS-1626008). The content is solely the responsibility of the authors and does not necessarily represent the official views of any of the funding agencies.
	
	
	
	\bibliography{bibfile}

\begin{thebibliography}{92}%
\makeatletter
\providecommand \@ifxundefined [1]{%
 \@ifx{#1\undefined}
}%
\providecommand \@ifnum [1]{%
 \ifnum #1\expandafter \@firstoftwo
 \else \expandafter \@secondoftwo
 \fi
}%
\providecommand \@ifx [1]{%
 \ifx #1\expandafter \@firstoftwo
 \else \expandafter \@secondoftwo
 \fi
}%
\providecommand \natexlab [1]{#1}%
\providecommand \enquote  [1]{``#1''}%
\providecommand \bibnamefont  [1]{#1}%
\providecommand \bibfnamefont [1]{#1}%
\providecommand \citenamefont [1]{#1}%
\providecommand \href@noop [0]{\@secondoftwo}%
\providecommand \href [0]{\begingroup \@sanitize@url \@href}%
\providecommand \@href[1]{\@@startlink{#1}\@@href}%
\providecommand \@@href[1]{\endgroup#1\@@endlink}%
\providecommand \@sanitize@url [0]{\catcode `\\12\catcode `\$12\catcode
  `\&12\catcode `\#12\catcode `\^12\catcode `\_12\catcode `\%12\relax}%
\providecommand \@@startlink[1]{}%
\providecommand \@@endlink[0]{}%
\providecommand \url  [0]{\begingroup\@sanitize@url \@url }%
\providecommand \@url [1]{\endgroup\@href {#1}{\urlprefix }}%
\providecommand \urlprefix  [0]{URL }%
\providecommand \Eprint [0]{\href }%
\providecommand \doibase [0]{http://dx.doi.org/}%
\providecommand \selectlanguage [0]{\@gobble}%
\providecommand \bibinfo  [0]{\@secondoftwo}%
\providecommand \bibfield  [0]{\@secondoftwo}%
\providecommand \translation [1]{[#1]}%
\providecommand \BibitemOpen [0]{}%
\providecommand \bibitemStop [0]{}%
\providecommand \bibitemNoStop [0]{.\EOS\space}%
\providecommand \EOS [0]{\spacefactor3000\relax}%
\providecommand \BibitemShut  [1]{\csname bibitem#1\endcsname}%
\let\auto@bib@innerbib\@empty
\bibitem [{\citenamefont {Zeki}\ and\ \citenamefont
  {Shipp}(1988)}]{zeki1988functional}%
  \BibitemOpen
  \bibfield  {author} {\bibinfo {author} {\bibfnamefont {Semir}\ \bibnamefont
  {Zeki}}\ and\ \bibinfo {author} {\bibfnamefont {Stewart}\ \bibnamefont
  {Shipp}},\ }\bibfield  {title} {\enquote {\bibinfo {title} {The functional
  logic of cortical connections},}\ }\href@noop {} {\bibfield  {journal}
  {\bibinfo  {journal} {Nature}\ }\textbf {\bibinfo {volume} {335}},\ \bibinfo
  {pages} {311--317} (\bibinfo {year} {1988})}\BibitemShut {NoStop}%
\bibitem [{\citenamefont {Sporns}\ \emph {et~al.}(2000)\citenamefont {Sporns},
  \citenamefont {Tononi},\ and\ \citenamefont
  {Edelman}}]{sporns2000theoretical}%
  \BibitemOpen
  \bibfield  {author} {\bibinfo {author} {\bibfnamefont {Olaf}\ \bibnamefont
  {Sporns}}, \bibinfo {author} {\bibfnamefont {Giulio}\ \bibnamefont {Tononi}},
  \ and\ \bibinfo {author} {\bibfnamefont {Gerald~M}\ \bibnamefont {Edelman}},\
  }\bibfield  {title} {\enquote {\bibinfo {title} {Theoretical neuroanatomy:
  relating anatomical and functional connectivity in graphs and cortical
  connection matrices},}\ }\href@noop {} {\bibfield  {journal} {\bibinfo
  {journal} {Cerebral cortex}\ }\textbf {\bibinfo {volume} {10}},\ \bibinfo
  {pages} {127--141} (\bibinfo {year} {2000})}\BibitemShut {NoStop}%
\bibitem [{\citenamefont {Hilgetag}\ \emph {et~al.}(2000)\citenamefont
  {Hilgetag}, \citenamefont {Burns}, \citenamefont {O'Neill}, \citenamefont
  {Scannell},\ and\ \citenamefont {Young}}]{hilgetag2000anatomical}%
  \BibitemOpen
  \bibfield  {author} {\bibinfo {author} {\bibfnamefont {Claus-C}\ \bibnamefont
  {Hilgetag}}, \bibinfo {author} {\bibfnamefont {Gully~APC}\ \bibnamefont
  {Burns}}, \bibinfo {author} {\bibfnamefont {Marc~A}\ \bibnamefont {O'Neill}},
  \bibinfo {author} {\bibfnamefont {Jack~W}\ \bibnamefont {Scannell}}, \ and\
  \bibinfo {author} {\bibfnamefont {Malcolm~P}\ \bibnamefont {Young}},\
  }\bibfield  {title} {\enquote {\bibinfo {title} {Anatomical connectivity
  defines the organization of clusters of cortical areas in the macaque and the
  cat},}\ }\href@noop {} {\bibfield  {journal} {\bibinfo  {journal}
  {Philosophical Transactions of the Royal Society of London B: Biological
  Sciences}\ }\textbf {\bibinfo {volume} {355}},\ \bibinfo {pages} {91--110}
  (\bibinfo {year} {2000})}\BibitemShut {NoStop}%
\bibitem [{\citenamefont {Stephan}\ \emph {et~al.}(2000)\citenamefont
  {Stephan}, \citenamefont {Hilgetag}, \citenamefont {Burns}, \citenamefont
  {O'Neill}, \citenamefont {Young},\ and\ \citenamefont
  {Kotter}}]{stephan2000computational}%
  \BibitemOpen
  \bibfield  {author} {\bibinfo {author} {\bibfnamefont {Klaas~E}\ \bibnamefont
  {Stephan}}, \bibinfo {author} {\bibfnamefont {Claus-C}\ \bibnamefont
  {Hilgetag}}, \bibinfo {author} {\bibfnamefont {Gully~APC}\ \bibnamefont
  {Burns}}, \bibinfo {author} {\bibfnamefont {Marc~A}\ \bibnamefont {O'Neill}},
  \bibinfo {author} {\bibfnamefont {Malcolm~P}\ \bibnamefont {Young}}, \ and\
  \bibinfo {author} {\bibfnamefont {Rolf}\ \bibnamefont {Kotter}},\ }\bibfield
  {title} {\enquote {\bibinfo {title} {Computational analysis of functional
  connectivity between areas of primate cerebral cortex},}\ }\href@noop {}
  {\bibfield  {journal} {\bibinfo  {journal} {Philosophical Transactions of the
  Royal Society of London B: Biological Sciences}\ }\textbf {\bibinfo {volume}
  {355}},\ \bibinfo {pages} {111--126} (\bibinfo {year} {2000})}\BibitemShut
  {NoStop}%
\bibitem [{\citenamefont {Passingham}\ \emph {et~al.}(2002)\citenamefont
  {Passingham}, \citenamefont {Stephan},\ and\ \citenamefont
  {K{\"o}tter}}]{passingham2002anatomical}%
  \BibitemOpen
  \bibfield  {author} {\bibinfo {author} {\bibfnamefont {Richard~E}\
  \bibnamefont {Passingham}}, \bibinfo {author} {\bibfnamefont {Klaas~E}\
  \bibnamefont {Stephan}}, \ and\ \bibinfo {author} {\bibfnamefont {Rolf}\
  \bibnamefont {K{\"o}tter}},\ }\bibfield  {title} {\enquote {\bibinfo {title}
  {The anatomical basis of functional localization in the cortex},}\
  }\href@noop {} {\bibfield  {journal} {\bibinfo  {journal} {Nature reviews.
  Neuroscience}\ }\textbf {\bibinfo {volume} {3}},\ \bibinfo {pages} {606}
  (\bibinfo {year} {2002})}\BibitemShut {NoStop}%
\bibitem [{\citenamefont {Sporns}\ \emph {et~al.}(2005)\citenamefont {Sporns},
  \citenamefont {Tononi},\ and\ \citenamefont {K{\"o}tter}}]{sporns2005human}%
  \BibitemOpen
  \bibfield  {author} {\bibinfo {author} {\bibfnamefont {Olaf}\ \bibnamefont
  {Sporns}}, \bibinfo {author} {\bibfnamefont {Giulio}\ \bibnamefont {Tononi}},
  \ and\ \bibinfo {author} {\bibfnamefont {Rolf}\ \bibnamefont {K{\"o}tter}},\
  }\bibfield  {title} {\enquote {\bibinfo {title} {The human connectome: a
  structural description of the human brain},}\ }\href@noop {} {\bibfield
  {journal} {\bibinfo  {journal} {PLoS computational biology}\ }\textbf
  {\bibinfo {volume} {1}},\ \bibinfo {pages} {e42} (\bibinfo {year}
  {2005})}\BibitemShut {NoStop}%
\bibitem [{\citenamefont {Rubinov}\ and\ \citenamefont
  {Sporns}(2010)}]{rubinov2010complex}%
  \BibitemOpen
  \bibfield  {author} {\bibinfo {author} {\bibfnamefont {Mikail}\ \bibnamefont
  {Rubinov}}\ and\ \bibinfo {author} {\bibfnamefont {Olaf}\ \bibnamefont
  {Sporns}},\ }\bibfield  {title} {\enquote {\bibinfo {title} {Complex network
  measures of brain connectivity: uses and interpretations},}\ }\href@noop {}
  {\bibfield  {journal} {\bibinfo  {journal} {Neuroimage}\ }\textbf {\bibinfo
  {volume} {52}},\ \bibinfo {pages} {1059--1069} (\bibinfo {year}
  {2010})}\BibitemShut {NoStop}%
\bibitem [{\citenamefont {Bassett}\ and\ \citenamefont
  {Sporns}(2017)}]{bassett2017network}%
  \BibitemOpen
  \bibfield  {author} {\bibinfo {author} {\bibfnamefont {Danielle~S}\
  \bibnamefont {Bassett}}\ and\ \bibinfo {author} {\bibfnamefont {Olaf}\
  \bibnamefont {Sporns}},\ }\bibfield  {title} {\enquote {\bibinfo {title}
  {Network neuroscience},}\ }\href@noop {} {\bibfield  {journal} {\bibinfo
  {journal} {Nature neuroscience}\ }\textbf {\bibinfo {volume} {20}},\ \bibinfo
  {pages} {353} (\bibinfo {year} {2017})}\BibitemShut {NoStop}%
\bibitem [{\citenamefont {Buzs{\'a}ki}\ \emph {et~al.}(2004)\citenamefont
  {Buzs{\'a}ki}, \citenamefont {Geisler}, \citenamefont {Henze},\ and\
  \citenamefont {Wang}}]{buzsaki2004interneuron}%
  \BibitemOpen
  \bibfield  {author} {\bibinfo {author} {\bibfnamefont {Gy{\"o}rgy}\
  \bibnamefont {Buzs{\'a}ki}}, \bibinfo {author} {\bibfnamefont {Caroline}\
  \bibnamefont {Geisler}}, \bibinfo {author} {\bibfnamefont {Darrell~A}\
  \bibnamefont {Henze}}, \ and\ \bibinfo {author} {\bibfnamefont {Xiao-Jing}\
  \bibnamefont {Wang}},\ }\bibfield  {title} {\enquote {\bibinfo {title}
  {Interneuron diversity series: circuit complexity and axon wiring economy of
  cortical interneurons},}\ }\href@noop {} {\bibfield  {journal} {\bibinfo
  {journal} {Trends in neurosciences}\ }\textbf {\bibinfo {volume} {27}},\
  \bibinfo {pages} {186--193} (\bibinfo {year} {2004})}\BibitemShut {NoStop}%
\bibitem [{\citenamefont {Chen}\ \emph {et~al.}(2006)\citenamefont {Chen},
  \citenamefont {Hall},\ and\ \citenamefont {Chklovskii}}]{chen2006wiring}%
  \BibitemOpen
  \bibfield  {author} {\bibinfo {author} {\bibfnamefont {Beth~L}\ \bibnamefont
  {Chen}}, \bibinfo {author} {\bibfnamefont {David~H}\ \bibnamefont {Hall}}, \
  and\ \bibinfo {author} {\bibfnamefont {Dmitri~B}\ \bibnamefont
  {Chklovskii}},\ }\bibfield  {title} {\enquote {\bibinfo {title} {Wiring
  optimization can relate neuronal structure and function},}\ }\href@noop {}
  {\bibfield  {journal} {\bibinfo  {journal} {Proceedings of the National
  Academy of Sciences of the United States of America}\ }\textbf {\bibinfo
  {volume} {103}},\ \bibinfo {pages} {4723--4728} (\bibinfo {year}
  {2006})}\BibitemShut {NoStop}%
\bibitem [{\citenamefont {Rivera-Alba}\ \emph {et~al.}(2011)\citenamefont
  {Rivera-Alba}, \citenamefont {Vitaladevuni}, \citenamefont {Mishchenko},
  \citenamefont {Lu}, \citenamefont {Takemura}, \citenamefont {Scheffer},
  \citenamefont {Meinertzhagen}, \citenamefont {Chklovskii},\ and\
  \citenamefont {de~Polavieja}}]{rivera2011wiring}%
  \BibitemOpen
  \bibfield  {author} {\bibinfo {author} {\bibfnamefont {Marta}\ \bibnamefont
  {Rivera-Alba}}, \bibinfo {author} {\bibfnamefont {Shiv~N}\ \bibnamefont
  {Vitaladevuni}}, \bibinfo {author} {\bibfnamefont {Yuriy}\ \bibnamefont
  {Mishchenko}}, \bibinfo {author} {\bibfnamefont {Zhiyuan}\ \bibnamefont
  {Lu}}, \bibinfo {author} {\bibfnamefont {Shin-ya}\ \bibnamefont {Takemura}},
  \bibinfo {author} {\bibfnamefont {Lou}\ \bibnamefont {Scheffer}}, \bibinfo
  {author} {\bibfnamefont {Ian~A}\ \bibnamefont {Meinertzhagen}}, \bibinfo
  {author} {\bibfnamefont {Dmitri~B}\ \bibnamefont {Chklovskii}}, \ and\
  \bibinfo {author} {\bibfnamefont {Gonzalo~G}\ \bibnamefont {de~Polavieja}},\
  }\bibfield  {title} {\enquote {\bibinfo {title} {Wiring economy and volume
  exclusion determine neuronal placement in the drosophila brain},}\
  }\href@noop {} {\bibfield  {journal} {\bibinfo  {journal} {Current Biology}\
  }\textbf {\bibinfo {volume} {21}},\ \bibinfo {pages} {2000--2005} (\bibinfo
  {year} {2011})}\BibitemShut {NoStop}%
\bibitem [{\citenamefont {Kaiser}\ and\ \citenamefont
  {Hilgetag}(2006)}]{kaiser2006nonoptimal}%
  \BibitemOpen
  \bibfield  {author} {\bibinfo {author} {\bibfnamefont {Marcus}\ \bibnamefont
  {Kaiser}}\ and\ \bibinfo {author} {\bibfnamefont {Claus~C}\ \bibnamefont
  {Hilgetag}},\ }\bibfield  {title} {\enquote {\bibinfo {title} {Nonoptimal
  component placement, but short processing paths, due to long-distance
  projections in neural systems},}\ }\href@noop {} {\bibfield  {journal}
  {\bibinfo  {journal} {PLoS computational biology}\ }\textbf {\bibinfo
  {volume} {2}},\ \bibinfo {pages} {e95} (\bibinfo {year} {2006})}\BibitemShut
  {NoStop}%
\bibitem [{\citenamefont {Horv{\'a}t}\ \emph {et~al.}(2016)\citenamefont
  {Horv{\'a}t}, \citenamefont {G{\u{a}}m{\u{a}}nut}, \citenamefont
  {Ercsey-Ravasz}, \citenamefont {Magrou}, \citenamefont {G{\u{a}}m{\u{a}}nut},
  \citenamefont {Van~Essen}, \citenamefont {Burkhalter}, \citenamefont
  {Knoblauch}, \citenamefont {Toroczkai},\ and\ \citenamefont
  {Kennedy}}]{horvat2016spatial}%
  \BibitemOpen
  \bibfield  {author} {\bibinfo {author} {\bibfnamefont {Szabolcs}\
  \bibnamefont {Horv{\'a}t}}, \bibinfo {author} {\bibfnamefont {R{\u{a}}zvan}\
  \bibnamefont {G{\u{a}}m{\u{a}}nut}}, \bibinfo {author} {\bibfnamefont
  {M{\'a}ria}\ \bibnamefont {Ercsey-Ravasz}}, \bibinfo {author} {\bibfnamefont
  {Lo{\"\i}c}\ \bibnamefont {Magrou}}, \bibinfo {author} {\bibfnamefont
  {Bianca}\ \bibnamefont {G{\u{a}}m{\u{a}}nut}}, \bibinfo {author}
  {\bibfnamefont {David~C}\ \bibnamefont {Van~Essen}}, \bibinfo {author}
  {\bibfnamefont {Andreas}\ \bibnamefont {Burkhalter}}, \bibinfo {author}
  {\bibfnamefont {Kenneth}\ \bibnamefont {Knoblauch}}, \bibinfo {author}
  {\bibfnamefont {Zolt{\'a}n}\ \bibnamefont {Toroczkai}}, \ and\ \bibinfo
  {author} {\bibfnamefont {Henry}\ \bibnamefont {Kennedy}},\ }\bibfield
  {title} {\enquote {\bibinfo {title} {Spatial embedding and wiring cost
  constrain the functional layout of the cortical network of rodents and
  primates},}\ }\href@noop {} {\bibfield  {journal} {\bibinfo  {journal} {PLoS
  biology}\ }\textbf {\bibinfo {volume} {14}},\ \bibinfo {pages} {e1002512}
  (\bibinfo {year} {2016})}\BibitemShut {NoStop}%
\bibitem [{\citenamefont {Rubinov}(2016)}]{rubinov2016constraints}%
  \BibitemOpen
  \bibfield  {author} {\bibinfo {author} {\bibfnamefont {Mikail}\ \bibnamefont
  {Rubinov}},\ }\bibfield  {title} {\enquote {\bibinfo {title} {Constraints and
  spandrels of interareal connectomes},}\ }\href@noop {} {\bibfield  {journal}
  {\bibinfo  {journal} {Nature communications}\ }\textbf {\bibinfo {volume}
  {7}},\ \bibinfo {pages} {13812} (\bibinfo {year} {2016})}\BibitemShut
  {NoStop}%
\bibitem [{\citenamefont {Raichle}\ and\ \citenamefont
  {Mintun}(2006)}]{raichle2006brain}%
  \BibitemOpen
  \bibfield  {author} {\bibinfo {author} {\bibfnamefont {Marcus~E}\
  \bibnamefont {Raichle}}\ and\ \bibinfo {author} {\bibfnamefont {Mark~A}\
  \bibnamefont {Mintun}},\ }\bibfield  {title} {\enquote {\bibinfo {title}
  {Brain work and brain imaging},}\ }\href@noop {} {\bibfield  {journal}
  {\bibinfo  {journal} {Annu. Rev. Neurosci.}\ }\textbf {\bibinfo {volume}
  {29}},\ \bibinfo {pages} {449--476} (\bibinfo {year} {2006})}\BibitemShut
  {NoStop}%
\bibitem [{\citenamefont {van~den Heuvel}\ \emph {et~al.}(2016)\citenamefont
  {van~den Heuvel}, \citenamefont {Bullmore},\ and\ \citenamefont
  {Sporns}}]{van2016comparative}%
  \BibitemOpen
  \bibfield  {author} {\bibinfo {author} {\bibfnamefont {Martijn~P}\
  \bibnamefont {van~den Heuvel}}, \bibinfo {author} {\bibfnamefont {Edward~T}\
  \bibnamefont {Bullmore}}, \ and\ \bibinfo {author} {\bibfnamefont {Olaf}\
  \bibnamefont {Sporns}},\ }\bibfield  {title} {\enquote {\bibinfo {title}
  {Comparative connectomics},}\ }\href@noop {} {\bibfield  {journal} {\bibinfo
  {journal} {Trends in cognitive sciences}\ }\textbf {\bibinfo {volume} {20}},\
  \bibinfo {pages} {345--361} (\bibinfo {year} {2016})}\BibitemShut {NoStop}%
\bibitem [{\citenamefont {Laughlin}\ and\ \citenamefont
  {Sejnowski}(2003)}]{laughlin2003communication}%
  \BibitemOpen
  \bibfield  {author} {\bibinfo {author} {\bibfnamefont {Simon~B}\ \bibnamefont
  {Laughlin}}\ and\ \bibinfo {author} {\bibfnamefont {Terrence~J}\ \bibnamefont
  {Sejnowski}},\ }\bibfield  {title} {\enquote {\bibinfo {title} {Communication
  in neuronal networks},}\ }\href@noop {} {\bibfield  {journal} {\bibinfo
  {journal} {Science}\ }\textbf {\bibinfo {volume} {301}},\ \bibinfo {pages}
  {1870--1874} (\bibinfo {year} {2003})}\BibitemShut {NoStop}%
\bibitem [{\citenamefont {Bullmore}\ and\ \citenamefont
  {Sporns}(2012)}]{bullmore2012economy}%
  \BibitemOpen
  \bibfield  {author} {\bibinfo {author} {\bibfnamefont {Ed}~\bibnamefont
  {Bullmore}}\ and\ \bibinfo {author} {\bibfnamefont {Olaf}\ \bibnamefont
  {Sporns}},\ }\bibfield  {title} {\enquote {\bibinfo {title} {The economy of
  brain network organization},}\ }\href@noop {} {\bibfield  {journal} {\bibinfo
   {journal} {Nature reviews. Neuroscience}\ }\textbf {\bibinfo {volume}
  {13}},\ \bibinfo {pages} {336} (\bibinfo {year} {2012})}\BibitemShut
  {NoStop}%
\bibitem [{\citenamefont {Chen}\ \emph {et~al.}(2013)\citenamefont {Chen},
  \citenamefont {Wang}, \citenamefont {Hilgetag},\ and\ \citenamefont
  {Zhou}}]{chen2013trade}%
  \BibitemOpen
  \bibfield  {author} {\bibinfo {author} {\bibfnamefont {Yuhan}\ \bibnamefont
  {Chen}}, \bibinfo {author} {\bibfnamefont {Shengjun}\ \bibnamefont {Wang}},
  \bibinfo {author} {\bibfnamefont {Claus~C}\ \bibnamefont {Hilgetag}}, \ and\
  \bibinfo {author} {\bibfnamefont {Changsong}\ \bibnamefont {Zhou}},\
  }\bibfield  {title} {\enquote {\bibinfo {title} {Trade-off between multiple
  constraints enables simultaneous formation of modules and hubs in neural
  systems},}\ }\href@noop {} {\bibfield  {journal} {\bibinfo  {journal} {PLoS
  computational biology}\ }\textbf {\bibinfo {volume} {9}},\ \bibinfo {pages}
  {e1002937} (\bibinfo {year} {2013})}\BibitemShut {NoStop}%
\bibitem [{\citenamefont {Sporns}\ and\ \citenamefont
  {Zwi}(2004)}]{sporns2004small}%
  \BibitemOpen
  \bibfield  {author} {\bibinfo {author} {\bibfnamefont {Olaf}\ \bibnamefont
  {Sporns}}\ and\ \bibinfo {author} {\bibfnamefont {Jonathan~D}\ \bibnamefont
  {Zwi}},\ }\bibfield  {title} {\enquote {\bibinfo {title} {The small world of
  the cerebral cortex},}\ }\href@noop {} {\bibfield  {journal} {\bibinfo
  {journal} {Neuroinformatics}\ }\textbf {\bibinfo {volume} {2}},\ \bibinfo
  {pages} {145--162} (\bibinfo {year} {2004})}\BibitemShut {NoStop}%
\bibitem [{\citenamefont {Van Den~Heuvel}\ and\ \citenamefont
  {Sporns}(2011)}]{van2011rich}%
  \BibitemOpen
  \bibfield  {author} {\bibinfo {author} {\bibfnamefont {Martijn~P}\
  \bibnamefont {Van Den~Heuvel}}\ and\ \bibinfo {author} {\bibfnamefont {Olaf}\
  \bibnamefont {Sporns}},\ }\bibfield  {title} {\enquote {\bibinfo {title}
  {Rich-club organization of the human connectome},}\ }\href@noop {} {\bibfield
   {journal} {\bibinfo  {journal} {Journal of Neuroscience}\ }\textbf {\bibinfo
  {volume} {31}},\ \bibinfo {pages} {15775--15786} (\bibinfo {year}
  {2011})}\BibitemShut {NoStop}%
\bibitem [{\citenamefont {Bassett}\ and\ \citenamefont
  {Bullmore}(2016)}]{bassett2016small}%
  \BibitemOpen
  \bibfield  {author} {\bibinfo {author} {\bibfnamefont {Danielle~S}\
  \bibnamefont {Bassett}}\ and\ \bibinfo {author} {\bibfnamefont {Edward~T}\
  \bibnamefont {Bullmore}},\ }\bibfield  {title} {\enquote {\bibinfo {title}
  {Small-world brain networks revisited},}\ }\href@noop {} {\bibfield
  {journal} {\bibinfo  {journal} {The Neuroscientist}\ ,\ \bibinfo {pages}
  {1073858416667720}} (\bibinfo {year} {2016})}\BibitemShut {NoStop}%
\bibitem [{\citenamefont {Ercsey-Ravasz}\ \emph {et~al.}(2013)\citenamefont
  {Ercsey-Ravasz}, \citenamefont {Markov}, \citenamefont {Lamy}, \citenamefont
  {Van~Essen}, \citenamefont {Knoblauch}, \citenamefont {Toroczkai},\ and\
  \citenamefont {Kennedy}}]{ercsey2013predictive}%
  \BibitemOpen
  \bibfield  {author} {\bibinfo {author} {\bibfnamefont {M{\'a}ria}\
  \bibnamefont {Ercsey-Ravasz}}, \bibinfo {author} {\bibfnamefont {Nikola~T}\
  \bibnamefont {Markov}}, \bibinfo {author} {\bibfnamefont {Camille}\
  \bibnamefont {Lamy}}, \bibinfo {author} {\bibfnamefont {David~C}\
  \bibnamefont {Van~Essen}}, \bibinfo {author} {\bibfnamefont {Kenneth}\
  \bibnamefont {Knoblauch}}, \bibinfo {author} {\bibfnamefont {Zolt{\'a}n}\
  \bibnamefont {Toroczkai}}, \ and\ \bibinfo {author} {\bibfnamefont {Henry}\
  \bibnamefont {Kennedy}},\ }\bibfield  {title} {\enquote {\bibinfo {title} {A
  predictive network model of cerebral cortical connectivity based on a
  distance rule},}\ }\href@noop {} {\bibfield  {journal} {\bibinfo  {journal}
  {Neuron}\ }\textbf {\bibinfo {volume} {80}},\ \bibinfo {pages} {184--197}
  (\bibinfo {year} {2013})}\BibitemShut {NoStop}%
\bibitem [{\citenamefont {Rubinov}\ \emph {et~al.}(2015)\citenamefont
  {Rubinov}, \citenamefont {Ypma}, \citenamefont {Watson},\ and\ \citenamefont
  {Bullmore}}]{rubinov2015wiring}%
  \BibitemOpen
  \bibfield  {author} {\bibinfo {author} {\bibfnamefont {Mikail}\ \bibnamefont
  {Rubinov}}, \bibinfo {author} {\bibfnamefont {Rolf~JF}\ \bibnamefont {Ypma}},
  \bibinfo {author} {\bibfnamefont {Charles}\ \bibnamefont {Watson}}, \ and\
  \bibinfo {author} {\bibfnamefont {Edward~T}\ \bibnamefont {Bullmore}},\
  }\bibfield  {title} {\enquote {\bibinfo {title} {Wiring cost and topological
  participation of the mouse brain connectome},}\ }\href@noop {} {\bibfield
  {journal} {\bibinfo  {journal} {Proceedings of the National Academy of
  Sciences}\ }\textbf {\bibinfo {volume} {112}},\ \bibinfo {pages}
  {10032--10037} (\bibinfo {year} {2015})}\BibitemShut {NoStop}%
\bibitem [{\citenamefont {Shih}\ \emph {et~al.}(2015)\citenamefont {Shih},
  \citenamefont {Sporns}, \citenamefont {Yuan}, \citenamefont {Su},
  \citenamefont {Lin}, \citenamefont {Chuang}, \citenamefont {Wang},
  \citenamefont {Lo}, \citenamefont {Greenspan},\ and\ \citenamefont
  {Chiang}}]{shih2015connectomics}%
  \BibitemOpen
  \bibfield  {author} {\bibinfo {author} {\bibfnamefont {Chi-Tin}\ \bibnamefont
  {Shih}}, \bibinfo {author} {\bibfnamefont {Olaf}\ \bibnamefont {Sporns}},
  \bibinfo {author} {\bibfnamefont {Shou-Li}\ \bibnamefont {Yuan}}, \bibinfo
  {author} {\bibfnamefont {Ta-Shun}\ \bibnamefont {Su}}, \bibinfo {author}
  {\bibfnamefont {Yen-Jen}\ \bibnamefont {Lin}}, \bibinfo {author}
  {\bibfnamefont {Chao-Chun}\ \bibnamefont {Chuang}}, \bibinfo {author}
  {\bibfnamefont {Ting-Yuan}\ \bibnamefont {Wang}}, \bibinfo {author}
  {\bibfnamefont {Chung-Chuang}\ \bibnamefont {Lo}}, \bibinfo {author}
  {\bibfnamefont {Ralph~J}\ \bibnamefont {Greenspan}}, \ and\ \bibinfo {author}
  {\bibfnamefont {Ann-Shyn}\ \bibnamefont {Chiang}},\ }\bibfield  {title}
  {\enquote {\bibinfo {title} {Connectomics-based analysis of information flow
  in the drosophila brain},}\ }\href@noop {} {\bibfield  {journal} {\bibinfo
  {journal} {Current Biology}\ }\textbf {\bibinfo {volume} {25}},\ \bibinfo
  {pages} {1249--1258} (\bibinfo {year} {2015})}\BibitemShut {NoStop}%
\bibitem [{\citenamefont {Roberts}\ \emph {et~al.}(2016)\citenamefont
  {Roberts}, \citenamefont {Perry}, \citenamefont {Lord}, \citenamefont
  {Roberts}, \citenamefont {Mitchell}, \citenamefont {Smith}, \citenamefont
  {Calamante},\ and\ \citenamefont {Breakspear}}]{roberts2016contribution}%
  \BibitemOpen
  \bibfield  {author} {\bibinfo {author} {\bibfnamefont {James~A}\ \bibnamefont
  {Roberts}}, \bibinfo {author} {\bibfnamefont {Alistair}\ \bibnamefont
  {Perry}}, \bibinfo {author} {\bibfnamefont {Anton~R}\ \bibnamefont {Lord}},
  \bibinfo {author} {\bibfnamefont {Gloria}\ \bibnamefont {Roberts}}, \bibinfo
  {author} {\bibfnamefont {Philip~B}\ \bibnamefont {Mitchell}}, \bibinfo
  {author} {\bibfnamefont {Robert~E}\ \bibnamefont {Smith}}, \bibinfo {author}
  {\bibfnamefont {Fernando}\ \bibnamefont {Calamante}}, \ and\ \bibinfo
  {author} {\bibfnamefont {Michael}\ \bibnamefont {Breakspear}},\ }\bibfield
  {title} {\enquote {\bibinfo {title} {The contribution of geometry to the
  human connectome},}\ }\href@noop {} {\bibfield  {journal} {\bibinfo
  {journal} {Neuroimage}\ }\textbf {\bibinfo {volume} {124}},\ \bibinfo {pages}
  {379--393} (\bibinfo {year} {2016})}\BibitemShut {NoStop}%
\bibitem [{\citenamefont {Mi{\v{s}}i{\'c}}\ \emph {et~al.}(2015)\citenamefont
  {Mi{\v{s}}i{\'c}}, \citenamefont {Betzel}, \citenamefont {Nematzadeh},
  \citenamefont {Go{\~n}i}, \citenamefont {Griffa}, \citenamefont {Hagmann},
  \citenamefont {Flammini}, \citenamefont {Ahn},\ and\ \citenamefont
  {Sporns}}]{mivsic2015cooperative}%
  \BibitemOpen
  \bibfield  {author} {\bibinfo {author} {\bibfnamefont {Bratislav}\
  \bibnamefont {Mi{\v{s}}i{\'c}}}, \bibinfo {author} {\bibfnamefont
  {Richard~F}\ \bibnamefont {Betzel}}, \bibinfo {author} {\bibfnamefont
  {Azadeh}\ \bibnamefont {Nematzadeh}}, \bibinfo {author} {\bibfnamefont
  {Joaquin}\ \bibnamefont {Go{\~n}i}}, \bibinfo {author} {\bibfnamefont
  {Alessandra}\ \bibnamefont {Griffa}}, \bibinfo {author} {\bibfnamefont
  {Patric}\ \bibnamefont {Hagmann}}, \bibinfo {author} {\bibfnamefont
  {Alessandro}\ \bibnamefont {Flammini}}, \bibinfo {author} {\bibfnamefont
  {Yong-Yeol}\ \bibnamefont {Ahn}}, \ and\ \bibinfo {author} {\bibfnamefont
  {Olaf}\ \bibnamefont {Sporns}},\ }\bibfield  {title} {\enquote {\bibinfo
  {title} {Cooperative and competitive spreading dynamics on the human
  connectome},}\ }\href@noop {} {\bibfield  {journal} {\bibinfo  {journal}
  {Neuron}\ }\textbf {\bibinfo {volume} {86}},\ \bibinfo {pages} {1518--1529}
  (\bibinfo {year} {2015})}\BibitemShut {NoStop}%
\bibitem [{\citenamefont {Avena-Koenigsberger}\ \emph
  {et~al.}(2017)\citenamefont {Avena-Koenigsberger}, \citenamefont
  {Mi{\v{s}}i{\'c}}, \citenamefont {Hawkins}, \citenamefont {Griffa},
  \citenamefont {Hagmann}, \citenamefont {Go{\~n}i},\ and\ \citenamefont
  {Sporns}}]{avena2017path}%
  \BibitemOpen
  \bibfield  {author} {\bibinfo {author} {\bibfnamefont {Andrea}\ \bibnamefont
  {Avena-Koenigsberger}}, \bibinfo {author} {\bibfnamefont {Bratislav}\
  \bibnamefont {Mi{\v{s}}i{\'c}}}, \bibinfo {author} {\bibfnamefont
  {Robert~XD}\ \bibnamefont {Hawkins}}, \bibinfo {author} {\bibfnamefont
  {Alessandra}\ \bibnamefont {Griffa}}, \bibinfo {author} {\bibfnamefont
  {Patric}\ \bibnamefont {Hagmann}}, \bibinfo {author} {\bibfnamefont
  {Joaqu{\'\i}n}\ \bibnamefont {Go{\~n}i}}, \ and\ \bibinfo {author}
  {\bibfnamefont {Olaf}\ \bibnamefont {Sporns}},\ }\bibfield  {title} {\enquote
  {\bibinfo {title} {Path ensembles and a tradeoff between communication
  efficiency and resilience in the human connectome},}\ }\href@noop {}
  {\bibfield  {journal} {\bibinfo  {journal} {Brain Structure and Function}\
  }\textbf {\bibinfo {volume} {222}},\ \bibinfo {pages} {603--618} (\bibinfo
  {year} {2017})}\BibitemShut {NoStop}%
\bibitem [{\citenamefont {Heiervang}\ \emph {et~al.}(2006)\citenamefont
  {Heiervang}, \citenamefont {Behrens}, \citenamefont {Mackay}, \citenamefont
  {Robson},\ and\ \citenamefont {Johansen-Berg}}]{heiervang2006between}%
  \BibitemOpen
  \bibfield  {author} {\bibinfo {author} {\bibfnamefont {E}~\bibnamefont
  {Heiervang}}, \bibinfo {author} {\bibfnamefont {TEJ}\ \bibnamefont
  {Behrens}}, \bibinfo {author} {\bibfnamefont {CE}~\bibnamefont {Mackay}},
  \bibinfo {author} {\bibfnamefont {MD}~\bibnamefont {Robson}}, \ and\ \bibinfo
  {author} {\bibfnamefont {H}~\bibnamefont {Johansen-Berg}},\ }\bibfield
  {title} {\enquote {\bibinfo {title} {Between session reproducibility and
  between subject variability of diffusion mr and tractography measures},}\
  }\href@noop {} {\bibfield  {journal} {\bibinfo  {journal} {Neuroimage}\
  }\textbf {\bibinfo {volume} {33}},\ \bibinfo {pages} {867--877} (\bibinfo
  {year} {2006})}\BibitemShut {NoStop}%
\bibitem [{\citenamefont {Bassett}\ \emph {et~al.}(2011)\citenamefont
  {Bassett}, \citenamefont {Brown}, \citenamefont {Deshpande}, \citenamefont
  {Carlson},\ and\ \citenamefont {Grafton}}]{bassett2011conserved}%
  \BibitemOpen
  \bibfield  {author} {\bibinfo {author} {\bibfnamefont {Danielle~S}\
  \bibnamefont {Bassett}}, \bibinfo {author} {\bibfnamefont {Jesse~A}\
  \bibnamefont {Brown}}, \bibinfo {author} {\bibfnamefont {Vibhas}\
  \bibnamefont {Deshpande}}, \bibinfo {author} {\bibfnamefont {Jean~M}\
  \bibnamefont {Carlson}}, \ and\ \bibinfo {author} {\bibfnamefont {Scott~T}\
  \bibnamefont {Grafton}},\ }\bibfield  {title} {\enquote {\bibinfo {title}
  {Conserved and variable architecture of human white matter connectivity},}\
  }\href@noop {} {\bibfield  {journal} {\bibinfo  {journal} {Neuroimage}\
  }\textbf {\bibinfo {volume} {54}},\ \bibinfo {pages} {1262--1279} (\bibinfo
  {year} {2011})}\BibitemShut {NoStop}%
\bibitem [{\citenamefont {Cammoun}\ \emph {et~al.}(2012)\citenamefont
  {Cammoun}, \citenamefont {Gigandet}, \citenamefont {Meskaldji}, \citenamefont
  {Thiran}, \citenamefont {Sporns}, \citenamefont {Do}, \citenamefont {Maeder},
  \citenamefont {Meuli},\ and\ \citenamefont {Hagmann}}]{cammoun2012mapping}%
  \BibitemOpen
  \bibfield  {author} {\bibinfo {author} {\bibfnamefont {Leila}\ \bibnamefont
  {Cammoun}}, \bibinfo {author} {\bibfnamefont {Xavier}\ \bibnamefont
  {Gigandet}}, \bibinfo {author} {\bibfnamefont {Djalel}\ \bibnamefont
  {Meskaldji}}, \bibinfo {author} {\bibfnamefont {Jean~Philippe}\ \bibnamefont
  {Thiran}}, \bibinfo {author} {\bibfnamefont {Olaf}\ \bibnamefont {Sporns}},
  \bibinfo {author} {\bibfnamefont {Kim~Q}\ \bibnamefont {Do}}, \bibinfo
  {author} {\bibfnamefont {Philippe}\ \bibnamefont {Maeder}}, \bibinfo {author}
  {\bibfnamefont {Reto}\ \bibnamefont {Meuli}}, \ and\ \bibinfo {author}
  {\bibfnamefont {Patric}\ \bibnamefont {Hagmann}},\ }\bibfield  {title}
  {\enquote {\bibinfo {title} {Mapping the human connectome at multiple scales
  with diffusion spectrum mri},}\ }\href@noop {} {\bibfield  {journal}
  {\bibinfo  {journal} {Journal of neuroscience methods}\ }\textbf {\bibinfo
  {volume} {203}},\ \bibinfo {pages} {386--397} (\bibinfo {year}
  {2012})}\BibitemShut {NoStop}%
\bibitem [{\citenamefont {Markov}\ \emph {et~al.}(2012)\citenamefont {Markov},
  \citenamefont {Ercsey-Ravasz}, \citenamefont {Ribeiro~Gomes}, \citenamefont
  {Lamy}, \citenamefont {Magrou}, \citenamefont {Vezoli}, \citenamefont
  {Misery}, \citenamefont {Falchier}, \citenamefont {Quilodran}, \citenamefont
  {Gariel} \emph {et~al.}}]{markov2012weighted}%
  \BibitemOpen
  \bibfield  {author} {\bibinfo {author} {\bibfnamefont {Nikola~T}\
  \bibnamefont {Markov}}, \bibinfo {author} {\bibfnamefont {MM}~\bibnamefont
  {Ercsey-Ravasz}}, \bibinfo {author} {\bibfnamefont {AR}~\bibnamefont
  {Ribeiro~Gomes}}, \bibinfo {author} {\bibfnamefont {Camille}\ \bibnamefont
  {Lamy}}, \bibinfo {author} {\bibfnamefont {Loic}\ \bibnamefont {Magrou}},
  \bibinfo {author} {\bibfnamefont {Julien}\ \bibnamefont {Vezoli}}, \bibinfo
  {author} {\bibfnamefont {P}~\bibnamefont {Misery}}, \bibinfo {author}
  {\bibfnamefont {A}~\bibnamefont {Falchier}}, \bibinfo {author} {\bibfnamefont
  {R}~\bibnamefont {Quilodran}}, \bibinfo {author} {\bibfnamefont
  {MA}~\bibnamefont {Gariel}},  \emph {et~al.},\ }\bibfield  {title} {\enquote
  {\bibinfo {title} {A weighted and directed interareal connectivity matrix for
  macaque cerebral cortex},}\ }\href@noop {} {\bibfield  {journal} {\bibinfo
  {journal} {Cerebral cortex}\ }\textbf {\bibinfo {volume} {24}},\ \bibinfo
  {pages} {17--36} (\bibinfo {year} {2012})}\BibitemShut {NoStop}%
\bibitem [{\citenamefont {Betzel}\ \emph
  {et~al.}(2017{\natexlab{a}})\citenamefont {Betzel}, \citenamefont
  {Medaglia},\ and\ \citenamefont {Bassett}}]{betzel2017diversity}%
  \BibitemOpen
  \bibfield  {author} {\bibinfo {author} {\bibfnamefont {Richard~F}\
  \bibnamefont {Betzel}}, \bibinfo {author} {\bibfnamefont {John~D}\
  \bibnamefont {Medaglia}}, \ and\ \bibinfo {author} {\bibfnamefont
  {Danielle~S}\ \bibnamefont {Bassett}},\ }\bibfield  {title} {\enquote
  {\bibinfo {title} {Diversity of meso-scale architecture in human and
  non-human connectomes},}\ }\href@noop {} {\bibfield  {journal} {\bibinfo
  {journal} {arXiv preprint arXiv:1702.02807}\ } (\bibinfo {year}
  {2017}{\natexlab{a}})}\BibitemShut {NoStop}%
\bibitem [{\citenamefont {Muldoon}\ \emph {et~al.}(2016)\citenamefont
  {Muldoon}, \citenamefont {Bridgeford},\ and\ \citenamefont
  {Bassett}}]{muldoon2016small}%
  \BibitemOpen
  \bibfield  {author} {\bibinfo {author} {\bibfnamefont {Sarah~Feldt}\
  \bibnamefont {Muldoon}}, \bibinfo {author} {\bibfnamefont {Eric~W}\
  \bibnamefont {Bridgeford}}, \ and\ \bibinfo {author} {\bibfnamefont
  {Danielle~S}\ \bibnamefont {Bassett}},\ }\bibfield  {title} {\enquote
  {\bibinfo {title} {Small-world propensity and weighted brain networks},}\
  }\href@noop {} {\bibfield  {journal} {\bibinfo  {journal} {Scientific
  reports}\ }\textbf {\bibinfo {volume} {6}},\ \bibinfo {pages} {22057}
  (\bibinfo {year} {2016})}\BibitemShut {NoStop}%
\bibitem [{\citenamefont {Go{\~n}i}\ \emph {et~al.}(2014)\citenamefont
  {Go{\~n}i}, \citenamefont {van~den Heuvel}, \citenamefont
  {Avena-Koenigsberger}, \citenamefont {de~Mendizabal}, \citenamefont {Betzel},
  \citenamefont {Griffa}, \citenamefont {Hagmann}, \citenamefont
  {Corominas-Murtra}, \citenamefont {Thiran},\ and\ \citenamefont
  {Sporns}}]{goni2014resting}%
  \BibitemOpen
  \bibfield  {author} {\bibinfo {author} {\bibfnamefont {Joaqu{\'\i}n}\
  \bibnamefont {Go{\~n}i}}, \bibinfo {author} {\bibfnamefont {Martijn~P}\
  \bibnamefont {van~den Heuvel}}, \bibinfo {author} {\bibfnamefont {Andrea}\
  \bibnamefont {Avena-Koenigsberger}}, \bibinfo {author} {\bibfnamefont
  {Nieves~Velez}\ \bibnamefont {de~Mendizabal}}, \bibinfo {author}
  {\bibfnamefont {Richard~F}\ \bibnamefont {Betzel}}, \bibinfo {author}
  {\bibfnamefont {Alessandra}\ \bibnamefont {Griffa}}, \bibinfo {author}
  {\bibfnamefont {Patric}\ \bibnamefont {Hagmann}}, \bibinfo {author}
  {\bibfnamefont {Bernat}\ \bibnamefont {Corominas-Murtra}}, \bibinfo {author}
  {\bibfnamefont {Jean-Philippe}\ \bibnamefont {Thiran}}, \ and\ \bibinfo
  {author} {\bibfnamefont {Olaf}\ \bibnamefont {Sporns}},\ }\bibfield  {title}
  {\enquote {\bibinfo {title} {Resting-brain functional connectivity predicted
  by analytic measures of network communication},}\ }\href@noop {} {\bibfield
  {journal} {\bibinfo  {journal} {Proceedings of the National Academy of
  Sciences}\ }\textbf {\bibinfo {volume} {111}},\ \bibinfo {pages} {833--838}
  (\bibinfo {year} {2014})}\BibitemShut {NoStop}%
\bibitem [{\citenamefont {Song}\ \emph {et~al.}(2014)\citenamefont {Song},
  \citenamefont {Kennedy},\ and\ \citenamefont {Wang}}]{song2014spatial}%
  \BibitemOpen
  \bibfield  {author} {\bibinfo {author} {\bibfnamefont {H~Francis}\
  \bibnamefont {Song}}, \bibinfo {author} {\bibfnamefont {Henry}\ \bibnamefont
  {Kennedy}}, \ and\ \bibinfo {author} {\bibfnamefont {Xiao-Jing}\ \bibnamefont
  {Wang}},\ }\bibfield  {title} {\enquote {\bibinfo {title} {Spatial embedding
  of structural similarity in the cerebral cortex},}\ }\href@noop {} {\bibfield
   {journal} {\bibinfo  {journal} {Proceedings of the National Academy of
  Sciences}\ }\textbf {\bibinfo {volume} {111}},\ \bibinfo {pages}
  {16580--16585} (\bibinfo {year} {2014})}\BibitemShut {NoStop}%
\bibitem [{\citenamefont {Gal{\'a}n}(2008)}]{galan2008network}%
  \BibitemOpen
  \bibfield  {author} {\bibinfo {author} {\bibfnamefont {Roberto~F}\
  \bibnamefont {Gal{\'a}n}},\ }\bibfield  {title} {\enquote {\bibinfo {title}
  {On how network architecture determines the dominant patterns of spontaneous
  neural activity},}\ }\href@noop {} {\bibfield  {journal} {\bibinfo  {journal}
  {PloS one}\ }\textbf {\bibinfo {volume} {3}},\ \bibinfo {pages} {e2148}
  (\bibinfo {year} {2008})}\BibitemShut {NoStop}%
\bibitem [{\citenamefont {K{\"o}tter}(2001)}]{kotter2001neuroscience}%
  \BibitemOpen
  \bibfield  {author} {\bibinfo {author} {\bibfnamefont {R}~\bibnamefont
  {K{\"o}tter}},\ }\bibfield  {title} {\enquote {\bibinfo {title} {Neuroscience
  databases: tools for exploring brain structure--function relationships},}\
  }\href@noop {} {\bibfield  {journal} {\bibinfo  {journal} {Philosophical
  Transactions of the Royal Society of London B: Biological Sciences}\ }\textbf
  {\bibinfo {volume} {356}},\ \bibinfo {pages} {1111--1120} (\bibinfo {year}
  {2001})}\BibitemShut {NoStop}%
\bibitem [{\citenamefont {Watts}\ and\ \citenamefont
  {Strogatz}(1998)}]{watts1998collective}%
  \BibitemOpen
  \bibfield  {author} {\bibinfo {author} {\bibfnamefont {Duncan~J}\
  \bibnamefont {Watts}}\ and\ \bibinfo {author} {\bibfnamefont {Steven~H}\
  \bibnamefont {Strogatz}},\ }\bibfield  {title} {\enquote {\bibinfo {title}
  {Collective dynamics of'small-world'networks},}\ }\href@noop {} {\bibfield
  {journal} {\bibinfo  {journal} {nature}\ }\textbf {\bibinfo {volume} {393}},\
  \bibinfo {pages} {440} (\bibinfo {year} {1998})}\BibitemShut {NoStop}%
\bibitem [{\citenamefont {Markov}\ \emph {et~al.}(2013)\citenamefont {Markov},
  \citenamefont {Ercsey-Ravasz}, \citenamefont {Lamy}, \citenamefont {Gomes},
  \citenamefont {Magrou}, \citenamefont {Misery}, \citenamefont {Giroud},
  \citenamefont {Barone}, \citenamefont {Dehay}, \citenamefont {Toroczkai}
  \emph {et~al.}}]{markov2013role}%
  \BibitemOpen
  \bibfield  {author} {\bibinfo {author} {\bibfnamefont {Nikola~T}\
  \bibnamefont {Markov}}, \bibinfo {author} {\bibfnamefont {Maria}\
  \bibnamefont {Ercsey-Ravasz}}, \bibinfo {author} {\bibfnamefont {Camille}\
  \bibnamefont {Lamy}}, \bibinfo {author} {\bibfnamefont {Ana Rita~Ribeiro}\
  \bibnamefont {Gomes}}, \bibinfo {author} {\bibfnamefont {Lo{\"\i}c}\
  \bibnamefont {Magrou}}, \bibinfo {author} {\bibfnamefont {Pierre}\
  \bibnamefont {Misery}}, \bibinfo {author} {\bibfnamefont {Pascale}\
  \bibnamefont {Giroud}}, \bibinfo {author} {\bibfnamefont {Pascal}\
  \bibnamefont {Barone}}, \bibinfo {author} {\bibfnamefont {Colette}\
  \bibnamefont {Dehay}}, \bibinfo {author} {\bibfnamefont {Zolt{\'a}n}\
  \bibnamefont {Toroczkai}},  \emph {et~al.},\ }\bibfield  {title} {\enquote
  {\bibinfo {title} {The role of long-range connections on the specificity of
  the macaque interareal cortical network},}\ }\href@noop {} {\bibfield
  {journal} {\bibinfo  {journal} {Proceedings of the National Academy of
  Sciences}\ }\textbf {\bibinfo {volume} {110}},\ \bibinfo {pages} {5187--5192}
  (\bibinfo {year} {2013})}\BibitemShut {NoStop}%
\bibitem [{\citenamefont {Chen}\ \emph {et~al.}(2017)\citenamefont {Chen},
  \citenamefont {Wang}, \citenamefont {Hilgetag},\ and\ \citenamefont
  {Zhou}}]{chen2017features}%
  \BibitemOpen
  \bibfield  {author} {\bibinfo {author} {\bibfnamefont {Yuhan}\ \bibnamefont
  {Chen}}, \bibinfo {author} {\bibfnamefont {Shengjun}\ \bibnamefont {Wang}},
  \bibinfo {author} {\bibfnamefont {Claus~C}\ \bibnamefont {Hilgetag}}, \ and\
  \bibinfo {author} {\bibfnamefont {Changsong}\ \bibnamefont {Zhou}},\
  }\bibfield  {title} {\enquote {\bibinfo {title} {Features of spatial and
  functional segregation and integration of the primate connectome revealed by
  trade-off between wiring cost and efficiency},}\ }\href@noop {} {\bibfield
  {journal} {\bibinfo  {journal} {PLOS Computational Biology}\ }\textbf
  {\bibinfo {volume} {13}},\ \bibinfo {pages} {e1005776} (\bibinfo {year}
  {2017})}\BibitemShut {NoStop}%
\bibitem [{\citenamefont {Samu}\ \emph {et~al.}(2014)\citenamefont {Samu},
  \citenamefont {Seth},\ and\ \citenamefont {Nowotny}}]{samu2014influence}%
  \BibitemOpen
  \bibfield  {author} {\bibinfo {author} {\bibfnamefont {David}\ \bibnamefont
  {Samu}}, \bibinfo {author} {\bibfnamefont {Anil~K}\ \bibnamefont {Seth}}, \
  and\ \bibinfo {author} {\bibfnamefont {Thomas}\ \bibnamefont {Nowotny}},\
  }\bibfield  {title} {\enquote {\bibinfo {title} {Influence of wiring cost on
  the large-scale architecture of human cortical connectivity},}\ }\href@noop
  {} {\bibfield  {journal} {\bibinfo  {journal} {PLoS computational biology}\
  }\textbf {\bibinfo {volume} {10}},\ \bibinfo {pages} {e1003557} (\bibinfo
  {year} {2014})}\BibitemShut {NoStop}%
\bibitem [{\citenamefont {Betzel}\ \emph
  {et~al.}(2017{\natexlab{b}})\citenamefont {Betzel}, \citenamefont {Medaglia},
  \citenamefont {Papadopoulos}, \citenamefont {Baum}, \citenamefont {Gur},
  \citenamefont {Gur}, \citenamefont {Roalf}, \citenamefont {Satterthwaite},\
  and\ \citenamefont {Bassett}}]{betzel2017modular}%
  \BibitemOpen
  \bibfield  {author} {\bibinfo {author} {\bibfnamefont {Richard~F}\
  \bibnamefont {Betzel}}, \bibinfo {author} {\bibfnamefont {John~D}\
  \bibnamefont {Medaglia}}, \bibinfo {author} {\bibfnamefont {Lia}\
  \bibnamefont {Papadopoulos}}, \bibinfo {author} {\bibfnamefont {Graham~L}\
  \bibnamefont {Baum}}, \bibinfo {author} {\bibfnamefont {Ruben}\ \bibnamefont
  {Gur}}, \bibinfo {author} {\bibfnamefont {Raquel}\ \bibnamefont {Gur}},
  \bibinfo {author} {\bibfnamefont {David}\ \bibnamefont {Roalf}}, \bibinfo
  {author} {\bibfnamefont {Theodore~D}\ \bibnamefont {Satterthwaite}}, \ and\
  \bibinfo {author} {\bibfnamefont {Danielle~S}\ \bibnamefont {Bassett}},\
  }\bibfield  {title} {\enquote {\bibinfo {title} {The modular organization of
  human anatomical brain networks: Accounting for the cost of wiring},}\
  }\href@noop {} {\bibfield  {journal} {\bibinfo  {journal} {Network
  Neuroscience}\ } (\bibinfo {year} {2017}{\natexlab{b}})}\BibitemShut
  {NoStop}%
\bibitem [{\citenamefont {Betzel}\ \emph {et~al.}(2016)\citenamefont {Betzel},
  \citenamefont {Avena-Koenigsberger}, \citenamefont {Go{\~n}i}, \citenamefont
  {He}, \citenamefont {De~Reus}, \citenamefont {Griffa}, \citenamefont
  {V{\'e}rtes}, \citenamefont {Mi{\v{s}}ic}, \citenamefont {Thiran},
  \citenamefont {Hagmann} \emph {et~al.}}]{betzel2016generative}%
  \BibitemOpen
  \bibfield  {author} {\bibinfo {author} {\bibfnamefont {Richard~F}\
  \bibnamefont {Betzel}}, \bibinfo {author} {\bibfnamefont {Andrea}\
  \bibnamefont {Avena-Koenigsberger}}, \bibinfo {author} {\bibfnamefont
  {Joaqu{\'\i}n}\ \bibnamefont {Go{\~n}i}}, \bibinfo {author} {\bibfnamefont
  {Ye}~\bibnamefont {He}}, \bibinfo {author} {\bibfnamefont {Marcel~A}\
  \bibnamefont {De~Reus}}, \bibinfo {author} {\bibfnamefont {Alessandra}\
  \bibnamefont {Griffa}}, \bibinfo {author} {\bibfnamefont {Petra~E}\
  \bibnamefont {V{\'e}rtes}}, \bibinfo {author} {\bibfnamefont {Bratislav}\
  \bibnamefont {Mi{\v{s}}ic}}, \bibinfo {author} {\bibfnamefont
  {Jean-Philippe}\ \bibnamefont {Thiran}}, \bibinfo {author} {\bibfnamefont
  {Patric}\ \bibnamefont {Hagmann}},  \emph {et~al.},\ }\bibfield  {title}
  {\enquote {\bibinfo {title} {Generative models of the human connectome},}\
  }\href@noop {} {\bibfield  {journal} {\bibinfo  {journal} {Neuroimage}\
  }\textbf {\bibinfo {volume} {124}},\ \bibinfo {pages} {1054--1064} (\bibinfo
  {year} {2016})}\BibitemShut {NoStop}%
\bibitem [{\citenamefont {Hilgetag}\ \emph {et~al.}(2016)\citenamefont
  {Hilgetag}, \citenamefont {Medalla}, \citenamefont {Beul},\ and\
  \citenamefont {Barbas}}]{hilgetag2016primate}%
  \BibitemOpen
  \bibfield  {author} {\bibinfo {author} {\bibfnamefont {Claus~C}\ \bibnamefont
  {Hilgetag}}, \bibinfo {author} {\bibfnamefont {Maria}\ \bibnamefont
  {Medalla}}, \bibinfo {author} {\bibfnamefont {Sarah~F}\ \bibnamefont {Beul}},
  \ and\ \bibinfo {author} {\bibfnamefont {Helen}\ \bibnamefont {Barbas}},\
  }\bibfield  {title} {\enquote {\bibinfo {title} {The primate connectome in
  context: principles of connections of the cortical visual system},}\
  }\href@noop {} {\bibfield  {journal} {\bibinfo  {journal} {NeuroImage}\
  }\textbf {\bibinfo {volume} {134}},\ \bibinfo {pages} {685--702} (\bibinfo
  {year} {2016})}\BibitemShut {NoStop}%
\bibitem [{\citenamefont {Beul}\ \emph {et~al.}(2015)\citenamefont {Beul},
  \citenamefont {Grant},\ and\ \citenamefont {Hilgetag}}]{beul2015predictive}%
  \BibitemOpen
  \bibfield  {author} {\bibinfo {author} {\bibfnamefont {Sarah~F}\ \bibnamefont
  {Beul}}, \bibinfo {author} {\bibfnamefont {Simon}\ \bibnamefont {Grant}}, \
  and\ \bibinfo {author} {\bibfnamefont {Claus~C}\ \bibnamefont {Hilgetag}},\
  }\bibfield  {title} {\enquote {\bibinfo {title} {A predictive model of the
  cat cortical connectome based on cytoarchitecture and distance},}\
  }\href@noop {} {\bibfield  {journal} {\bibinfo  {journal} {Brain Structure
  and Function}\ }\textbf {\bibinfo {volume} {220}},\ \bibinfo {pages}
  {3167--3184} (\bibinfo {year} {2015})}\BibitemShut {NoStop}%
\bibitem [{\citenamefont {Richiardi}\ \emph {et~al.}(2015)\citenamefont
  {Richiardi}, \citenamefont {Altmann}, \citenamefont {Milazzo}, \citenamefont
  {Chang}, \citenamefont {Chakravarty}, \citenamefont {Banaschewski},
  \citenamefont {Barker}, \citenamefont {Bokde}, \citenamefont {Bromberg},
  \citenamefont {B{\"u}chel} \emph {et~al.}}]{richiardi2015correlated}%
  \BibitemOpen
  \bibfield  {author} {\bibinfo {author} {\bibfnamefont {Jonas}\ \bibnamefont
  {Richiardi}}, \bibinfo {author} {\bibfnamefont {Andre}\ \bibnamefont
  {Altmann}}, \bibinfo {author} {\bibfnamefont {Anna-Clare}\ \bibnamefont
  {Milazzo}}, \bibinfo {author} {\bibfnamefont {Catie}\ \bibnamefont {Chang}},
  \bibinfo {author} {\bibfnamefont {M~Mallar}\ \bibnamefont {Chakravarty}},
  \bibinfo {author} {\bibfnamefont {Tobias}\ \bibnamefont {Banaschewski}},
  \bibinfo {author} {\bibfnamefont {Gareth~J}\ \bibnamefont {Barker}}, \bibinfo
  {author} {\bibfnamefont {Arun~LW}\ \bibnamefont {Bokde}}, \bibinfo {author}
  {\bibfnamefont {Uli}\ \bibnamefont {Bromberg}}, \bibinfo {author}
  {\bibfnamefont {Christian}\ \bibnamefont {B{\"u}chel}},  \emph {et~al.},\
  }\bibfield  {title} {\enquote {\bibinfo {title} {Correlated gene expression
  supports synchronous activity in brain networks},}\ }\href@noop {} {\bibfield
   {journal} {\bibinfo  {journal} {Science}\ }\textbf {\bibinfo {volume}
  {348}},\ \bibinfo {pages} {1241--1244} (\bibinfo {year} {2015})}\BibitemShut
  {NoStop}%
\bibitem [{\citenamefont {Henriksen}\ \emph {et~al.}(2016)\citenamefont
  {Henriksen}, \citenamefont {Pang},\ and\ \citenamefont
  {Wronkiewicz}}]{henriksen2016simple}%
  \BibitemOpen
  \bibfield  {author} {\bibinfo {author} {\bibfnamefont {Sid}\ \bibnamefont
  {Henriksen}}, \bibinfo {author} {\bibfnamefont {Rich}\ \bibnamefont {Pang}},
  \ and\ \bibinfo {author} {\bibfnamefont {Mark}\ \bibnamefont {Wronkiewicz}},\
  }\bibfield  {title} {\enquote {\bibinfo {title} {A simple generative model of
  the mouse mesoscale connectome},}\ }\href@noop {} {\bibfield  {journal}
  {\bibinfo  {journal} {Elife}\ }\textbf {\bibinfo {volume} {5}},\ \bibinfo
  {pages} {e12366} (\bibinfo {year} {2016})}\BibitemShut {NoStop}%
\bibitem [{\citenamefont {Kaiser}(2017)}]{kaiser2017mechanisms}%
  \BibitemOpen
  \bibfield  {author} {\bibinfo {author} {\bibfnamefont {Marcus}\ \bibnamefont
  {Kaiser}},\ }\bibfield  {title} {\enquote {\bibinfo {title} {Mechanisms of
  connectome development},}\ }\href@noop {} {\bibfield  {journal} {\bibinfo
  {journal} {Trends in Cognitive Sciences}\ } (\bibinfo {year}
  {2017})}\BibitemShut {NoStop}%
\bibitem [{\citenamefont {V{\'e}rtes}\ \emph {et~al.}(2012)\citenamefont
  {V{\'e}rtes}, \citenamefont {Alexander-Bloch}, \citenamefont {Gogtay},
  \citenamefont {Giedd}, \citenamefont {Rapoport},\ and\ \citenamefont
  {Bullmore}}]{vertes2012simple}%
  \BibitemOpen
  \bibfield  {author} {\bibinfo {author} {\bibfnamefont {Petra~E}\ \bibnamefont
  {V{\'e}rtes}}, \bibinfo {author} {\bibfnamefont {Aaron~F}\ \bibnamefont
  {Alexander-Bloch}}, \bibinfo {author} {\bibfnamefont {Nitin}\ \bibnamefont
  {Gogtay}}, \bibinfo {author} {\bibfnamefont {Jay~N}\ \bibnamefont {Giedd}},
  \bibinfo {author} {\bibfnamefont {Judith~L}\ \bibnamefont {Rapoport}}, \ and\
  \bibinfo {author} {\bibfnamefont {Edward~T}\ \bibnamefont {Bullmore}},\
  }\bibfield  {title} {\enquote {\bibinfo {title} {Simple models of human brain
  functional networks},}\ }\href@noop {} {\bibfield  {journal} {\bibinfo
  {journal} {Proceedings of the National Academy of Sciences}\ }\textbf
  {\bibinfo {volume} {109}},\ \bibinfo {pages} {5868--5873} (\bibinfo {year}
  {2012})}\BibitemShut {NoStop}%
\bibitem [{\citenamefont {Betzel}\ and\ \citenamefont
  {Bassett}(2017)}]{betzel2017generative}%
  \BibitemOpen
  \bibfield  {author} {\bibinfo {author} {\bibfnamefont {Richard~F}\
  \bibnamefont {Betzel}}\ and\ \bibinfo {author} {\bibfnamefont {Danielle~S}\
  \bibnamefont {Bassett}},\ }\bibfield  {title} {\enquote {\bibinfo {title}
  {Generative models for network neuroscience: Prospects and promise},}\
  }\href@noop {} {\bibfield  {journal} {\bibinfo  {journal} {arXiv preprint
  arXiv:1708.07958}\ } (\bibinfo {year} {2017})}\BibitemShut {NoStop}%
\bibitem [{\citenamefont {Hursh}(1939)}]{hursh1939conduction}%
  \BibitemOpen
  \bibfield  {author} {\bibinfo {author} {\bibfnamefont {JB}~\bibnamefont
  {Hursh}},\ }\bibfield  {title} {\enquote {\bibinfo {title} {Conduction
  velocity and diameter of nerve fibers},}\ }\href@noop {} {\bibfield
  {journal} {\bibinfo  {journal} {American Journal of Physiology--Legacy
  Content}\ }\textbf {\bibinfo {volume} {127}},\ \bibinfo {pages} {131--139}
  (\bibinfo {year} {1939})}\BibitemShut {NoStop}%
\bibitem [{\citenamefont {Ritchie}(1982)}]{ritchie1982relation}%
  \BibitemOpen
  \bibfield  {author} {\bibinfo {author} {\bibfnamefont {JM}~\bibnamefont
  {Ritchie}},\ }\bibfield  {title} {\enquote {\bibinfo {title} {On the relation
  between fibre diameter and conduction velocity in myelinated nerve fibres},}\
  }\href@noop {} {\bibfield  {journal} {\bibinfo  {journal} {Proceedings of the
  Royal Society of London B: Biological Sciences}\ }\textbf {\bibinfo {volume}
  {217}},\ \bibinfo {pages} {29--35} (\bibinfo {year} {1982})}\BibitemShut
  {NoStop}%
\bibitem [{\citenamefont {Buzs{\'a}ki}\ and\ \citenamefont
  {Mizuseki}(2014)}]{buzsaki2014log}%
  \BibitemOpen
  \bibfield  {author} {\bibinfo {author} {\bibfnamefont {Gy{\"o}rgy}\
  \bibnamefont {Buzs{\'a}ki}}\ and\ \bibinfo {author} {\bibfnamefont {Kenji}\
  \bibnamefont {Mizuseki}},\ }\bibfield  {title} {\enquote {\bibinfo {title}
  {The log-dynamic brain: how skewed distributions affect network
  operations},}\ }\href@noop {} {\bibfield  {journal} {\bibinfo  {journal}
  {Nature reviews. Neuroscience}\ }\textbf {\bibinfo {volume} {15}},\ \bibinfo
  {pages} {264} (\bibinfo {year} {2014})}\BibitemShut {NoStop}%
\bibitem [{\citenamefont {Markov}\ \emph {et~al.}(2010)\citenamefont {Markov},
  \citenamefont {Misery}, \citenamefont {Falchier}, \citenamefont {Lamy},
  \citenamefont {Vezoli}, \citenamefont {Quilodran}, \citenamefont {Gariel},
  \citenamefont {Giroud}, \citenamefont {Ercsey-Ravasz}, \citenamefont {Pilaz}
  \emph {et~al.}}]{markov2010weight}%
  \BibitemOpen
  \bibfield  {author} {\bibinfo {author} {\bibfnamefont {NT}~\bibnamefont
  {Markov}}, \bibinfo {author} {\bibfnamefont {P}~\bibnamefont {Misery}},
  \bibinfo {author} {\bibfnamefont {A}~\bibnamefont {Falchier}}, \bibinfo
  {author} {\bibfnamefont {C}~\bibnamefont {Lamy}}, \bibinfo {author}
  {\bibfnamefont {J}~\bibnamefont {Vezoli}}, \bibinfo {author} {\bibfnamefont
  {R}~\bibnamefont {Quilodran}}, \bibinfo {author} {\bibfnamefont
  {MA}~\bibnamefont {Gariel}}, \bibinfo {author} {\bibfnamefont
  {P}~\bibnamefont {Giroud}}, \bibinfo {author} {\bibfnamefont {M}~\bibnamefont
  {Ercsey-Ravasz}}, \bibinfo {author} {\bibfnamefont {LJ}~\bibnamefont
  {Pilaz}},  \emph {et~al.},\ }\bibfield  {title} {\enquote {\bibinfo {title}
  {Weight consistency specifies regularities of macaque cortical networks},}\
  }\href@noop {} {\bibfield  {journal} {\bibinfo  {journal} {Cerebral Cortex}\
  }\textbf {\bibinfo {volume} {21}},\ \bibinfo {pages} {1254--1272} (\bibinfo
  {year} {2010})}\BibitemShut {NoStop}%
\bibitem [{\citenamefont {Deco}\ \emph {et~al.}(2008)\citenamefont {Deco},
  \citenamefont {Jirsa}, \citenamefont {Robinson}, \citenamefont {Breakspear},\
  and\ \citenamefont {Friston}}]{deco2008dynamic}%
  \BibitemOpen
  \bibfield  {author} {\bibinfo {author} {\bibfnamefont {Gustavo}\ \bibnamefont
  {Deco}}, \bibinfo {author} {\bibfnamefont {Viktor~K}\ \bibnamefont {Jirsa}},
  \bibinfo {author} {\bibfnamefont {Peter~A}\ \bibnamefont {Robinson}},
  \bibinfo {author} {\bibfnamefont {Michael}\ \bibnamefont {Breakspear}}, \
  and\ \bibinfo {author} {\bibfnamefont {Karl}\ \bibnamefont {Friston}},\
  }\bibfield  {title} {\enquote {\bibinfo {title} {The dynamic brain: from
  spiking neurons to neural masses and cortical fields},}\ }\href@noop {}
  {\bibfield  {journal} {\bibinfo  {journal} {PLoS computational biology}\
  }\textbf {\bibinfo {volume} {4}},\ \bibinfo {pages} {e1000092} (\bibinfo
  {year} {2008})}\BibitemShut {NoStop}%
\bibitem [{\citenamefont {Leon}\ \emph {et~al.}(2013)\citenamefont {Leon},
  \citenamefont {Knock}, \citenamefont {Woodman}, \citenamefont {Domide},
  \citenamefont {Mersmann}, \citenamefont {McIntosh},\ and\ \citenamefont
  {Jirsa}}]{leon2013virtual}%
  \BibitemOpen
  \bibfield  {author} {\bibinfo {author} {\bibfnamefont {Paula~Sanz}\
  \bibnamefont {Leon}}, \bibinfo {author} {\bibfnamefont {Stuart~A}\
  \bibnamefont {Knock}}, \bibinfo {author} {\bibfnamefont {M~Marmaduke}\
  \bibnamefont {Woodman}}, \bibinfo {author} {\bibfnamefont {Lia}\ \bibnamefont
  {Domide}}, \bibinfo {author} {\bibfnamefont {Jochen}\ \bibnamefont
  {Mersmann}}, \bibinfo {author} {\bibfnamefont {Anthony~R}\ \bibnamefont
  {McIntosh}}, \ and\ \bibinfo {author} {\bibfnamefont {Viktor}\ \bibnamefont
  {Jirsa}},\ }\bibfield  {title} {\enquote {\bibinfo {title} {The virtual
  brain: a simulator of primate brain network dynamics},}\ }\href@noop {}
  {\bibfield  {journal} {\bibinfo  {journal} {Frontiers in neuroinformatics}\
  }\textbf {\bibinfo {volume} {7}} (\bibinfo {year} {2013})}\BibitemShut
  {NoStop}%
\bibitem [{\citenamefont {Chaudhuri}\ \emph {et~al.}(2015)\citenamefont
  {Chaudhuri}, \citenamefont {Knoblauch}, \citenamefont {Gariel}, \citenamefont
  {Kennedy},\ and\ \citenamefont {Wang}}]{chaudhuri2015large}%
  \BibitemOpen
  \bibfield  {author} {\bibinfo {author} {\bibfnamefont {Rishidev}\
  \bibnamefont {Chaudhuri}}, \bibinfo {author} {\bibfnamefont {Kenneth}\
  \bibnamefont {Knoblauch}}, \bibinfo {author} {\bibfnamefont {Marie-Alice}\
  \bibnamefont {Gariel}}, \bibinfo {author} {\bibfnamefont {Henry}\
  \bibnamefont {Kennedy}}, \ and\ \bibinfo {author} {\bibfnamefont {Xiao-Jing}\
  \bibnamefont {Wang}},\ }\bibfield  {title} {\enquote {\bibinfo {title} {A
  large-scale circuit mechanism for hierarchical dynamical processing in the
  primate cortex},}\ }\href@noop {} {\bibfield  {journal} {\bibinfo  {journal}
  {Neuron}\ }\textbf {\bibinfo {volume} {88}},\ \bibinfo {pages} {419--431}
  (\bibinfo {year} {2015})}\BibitemShut {NoStop}%
\bibitem [{\citenamefont {Honey}\ \emph {et~al.}(2009)\citenamefont {Honey},
  \citenamefont {Sporns}, \citenamefont {Cammoun}, \citenamefont {Gigandet},
  \citenamefont {Thiran}, \citenamefont {Meuli},\ and\ \citenamefont
  {Hagmann}}]{honey2009predicting}%
  \BibitemOpen
  \bibfield  {author} {\bibinfo {author} {\bibfnamefont {CJ}~\bibnamefont
  {Honey}}, \bibinfo {author} {\bibfnamefont {O}~\bibnamefont {Sporns}},
  \bibinfo {author} {\bibfnamefont {Leila}\ \bibnamefont {Cammoun}}, \bibinfo
  {author} {\bibfnamefont {Xavier}\ \bibnamefont {Gigandet}}, \bibinfo {author}
  {\bibfnamefont {Jean-Philippe}\ \bibnamefont {Thiran}}, \bibinfo {author}
  {\bibfnamefont {Reto}\ \bibnamefont {Meuli}}, \ and\ \bibinfo {author}
  {\bibfnamefont {Patric}\ \bibnamefont {Hagmann}},\ }\bibfield  {title}
  {\enquote {\bibinfo {title} {Predicting human resting-state functional
  connectivity from structural connectivity},}\ }\href@noop {} {\bibfield
  {journal} {\bibinfo  {journal} {Proceedings of the National Academy of
  Sciences}\ }\textbf {\bibinfo {volume} {106}},\ \bibinfo {pages} {2035--2040}
  (\bibinfo {year} {2009})}\BibitemShut {NoStop}%
\bibitem [{\citenamefont {Abdelnour}\ \emph {et~al.}(2014)\citenamefont
  {Abdelnour}, \citenamefont {Voss},\ and\ \citenamefont
  {Raj}}]{abdelnour2014network}%
  \BibitemOpen
  \bibfield  {author} {\bibinfo {author} {\bibfnamefont {Farras}\ \bibnamefont
  {Abdelnour}}, \bibinfo {author} {\bibfnamefont {Henning~U}\ \bibnamefont
  {Voss}}, \ and\ \bibinfo {author} {\bibfnamefont {Ashish}\ \bibnamefont
  {Raj}},\ }\bibfield  {title} {\enquote {\bibinfo {title} {Network diffusion
  accurately models the relationship between structural and functional brain
  connectivity networks},}\ }\href@noop {} {\bibfield  {journal} {\bibinfo
  {journal} {Neuroimage}\ }\textbf {\bibinfo {volume} {90}},\ \bibinfo {pages}
  {335--347} (\bibinfo {year} {2014})}\BibitemShut {NoStop}%
\bibitem [{\citenamefont {Becker}\ \emph {et~al.}(2015)\citenamefont {Becker},
  \citenamefont {Pequito}, \citenamefont {Pappas}, \citenamefont {Miller},
  \citenamefont {Grafton}, \citenamefont {Bassett},\ and\ \citenamefont
  {Preciado}}]{becker2015accurately}%
  \BibitemOpen
  \bibfield  {author} {\bibinfo {author} {\bibfnamefont {Cassiano~O}\
  \bibnamefont {Becker}}, \bibinfo {author} {\bibfnamefont {Sergio}\
  \bibnamefont {Pequito}}, \bibinfo {author} {\bibfnamefont {George~J}\
  \bibnamefont {Pappas}}, \bibinfo {author} {\bibfnamefont {Michael~B}\
  \bibnamefont {Miller}}, \bibinfo {author} {\bibfnamefont {Scott~T}\
  \bibnamefont {Grafton}}, \bibinfo {author} {\bibfnamefont {Danielle~S}\
  \bibnamefont {Bassett}}, \ and\ \bibinfo {author} {\bibfnamefont {Victor~M}\
  \bibnamefont {Preciado}},\ }\bibfield  {title} {\enquote {\bibinfo {title}
  {Accurately predicting functional connectivity from diffusion imaging},}\
  }\href@noop {} {\bibfield  {journal} {\bibinfo  {journal} {arXiv preprint
  arXiv:1512.02602}\ } (\bibinfo {year} {2015})}\BibitemShut {NoStop}%
\bibitem [{\citenamefont {Crossley}\ \emph {et~al.}(2014)\citenamefont
  {Crossley}, \citenamefont {Mechelli}, \citenamefont {Scott}, \citenamefont
  {Carletti}, \citenamefont {Fox}, \citenamefont {McGuire},\ and\ \citenamefont
  {Bullmore}}]{crossley2014hubs}%
  \BibitemOpen
  \bibfield  {author} {\bibinfo {author} {\bibfnamefont {Nicolas~A}\
  \bibnamefont {Crossley}}, \bibinfo {author} {\bibfnamefont {Andrea}\
  \bibnamefont {Mechelli}}, \bibinfo {author} {\bibfnamefont {Jessica}\
  \bibnamefont {Scott}}, \bibinfo {author} {\bibfnamefont {Francesco}\
  \bibnamefont {Carletti}}, \bibinfo {author} {\bibfnamefont {Peter~T}\
  \bibnamefont {Fox}}, \bibinfo {author} {\bibfnamefont {Philip}\ \bibnamefont
  {McGuire}}, \ and\ \bibinfo {author} {\bibfnamefont {Edward~T}\ \bibnamefont
  {Bullmore}},\ }\bibfield  {title} {\enquote {\bibinfo {title} {The hubs of
  the human connectome are generally implicated in the anatomy of brain
  disorders},}\ }\href@noop {} {\bibfield  {journal} {\bibinfo  {journal}
  {Brain}\ }\textbf {\bibinfo {volume} {137}},\ \bibinfo {pages} {2382--2395}
  (\bibinfo {year} {2014})}\BibitemShut {NoStop}%
\bibitem [{\citenamefont {Korgaonkar}\ \emph {et~al.}(2014)\citenamefont
  {Korgaonkar}, \citenamefont {Fornito}, \citenamefont {Williams},\ and\
  \citenamefont {Grieve}}]{korgaonkar2014abnormal}%
  \BibitemOpen
  \bibfield  {author} {\bibinfo {author} {\bibfnamefont {Mayuresh~S}\
  \bibnamefont {Korgaonkar}}, \bibinfo {author} {\bibfnamefont {Alex}\
  \bibnamefont {Fornito}}, \bibinfo {author} {\bibfnamefont {Leanne~M}\
  \bibnamefont {Williams}}, \ and\ \bibinfo {author} {\bibfnamefont {Stuart~M}\
  \bibnamefont {Grieve}},\ }\bibfield  {title} {\enquote {\bibinfo {title}
  {Abnormal structural networks characterize major depressive disorder: a
  connectome analysis},}\ }\href@noop {} {\bibfield  {journal} {\bibinfo
  {journal} {Biological psychiatry}\ }\textbf {\bibinfo {volume} {76}},\
  \bibinfo {pages} {567--574} (\bibinfo {year} {2014})}\BibitemShut {NoStop}%
\bibitem [{\citenamefont {Reveley}\ \emph {et~al.}(2015)\citenamefont
  {Reveley}, \citenamefont {Seth}, \citenamefont {Pierpaoli}, \citenamefont
  {Silva}, \citenamefont {Yu}, \citenamefont {Saunders}, \citenamefont
  {Leopold},\ and\ \citenamefont {Frank}}]{reveley2015superficial}%
  \BibitemOpen
  \bibfield  {author} {\bibinfo {author} {\bibfnamefont {Colin}\ \bibnamefont
  {Reveley}}, \bibinfo {author} {\bibfnamefont {Anil~K}\ \bibnamefont {Seth}},
  \bibinfo {author} {\bibfnamefont {Carlo}\ \bibnamefont {Pierpaoli}}, \bibinfo
  {author} {\bibfnamefont {Afonso~C}\ \bibnamefont {Silva}}, \bibinfo {author}
  {\bibfnamefont {David}\ \bibnamefont {Yu}}, \bibinfo {author} {\bibfnamefont
  {Richard~C}\ \bibnamefont {Saunders}}, \bibinfo {author} {\bibfnamefont
  {David~A}\ \bibnamefont {Leopold}}, \ and\ \bibinfo {author} {\bibfnamefont
  {Q~Ye}\ \bibnamefont {Frank}},\ }\bibfield  {title} {\enquote {\bibinfo
  {title} {Superficial white matter fiber systems impede detection of
  long-range cortical connections in diffusion mr tractography},}\ }\href@noop
  {} {\bibfield  {journal} {\bibinfo  {journal} {Proceedings of the National
  Academy of Sciences}\ }\textbf {\bibinfo {volume} {112}},\ \bibinfo {pages}
  {E2820--E2828} (\bibinfo {year} {2015})}\BibitemShut {NoStop}%
\bibitem [{\citenamefont {Sotiropoulos}\ and\ \citenamefont
  {Zalesky}(2017)}]{sotiropoulos2017building}%
  \BibitemOpen
  \bibfield  {author} {\bibinfo {author} {\bibfnamefont {Stamatios~N}\
  \bibnamefont {Sotiropoulos}}\ and\ \bibinfo {author} {\bibfnamefont {Andrew}\
  \bibnamefont {Zalesky}},\ }\bibfield  {title} {\enquote {\bibinfo {title}
  {Building connectomes using diffusion mri: Why, how and but},}\ }\href@noop
  {} {\bibfield  {journal} {\bibinfo  {journal} {NMR in Biomedicine}\ }
  (\bibinfo {year} {2017})}\BibitemShut {NoStop}%
\bibitem [{\citenamefont {Pestilli}\ \emph {et~al.}(2014)\citenamefont
  {Pestilli}, \citenamefont {Yeatman}, \citenamefont {Rokem}, \citenamefont
  {Kay},\ and\ \citenamefont {Wandell}}]{pestilli2014evaluation}%
  \BibitemOpen
  \bibfield  {author} {\bibinfo {author} {\bibfnamefont {Franco}\ \bibnamefont
  {Pestilli}}, \bibinfo {author} {\bibfnamefont {Jason~D}\ \bibnamefont
  {Yeatman}}, \bibinfo {author} {\bibfnamefont {Ariel}\ \bibnamefont {Rokem}},
  \bibinfo {author} {\bibfnamefont {Kendrick~N}\ \bibnamefont {Kay}}, \ and\
  \bibinfo {author} {\bibfnamefont {Brian~A}\ \bibnamefont {Wandell}},\
  }\bibfield  {title} {\enquote {\bibinfo {title} {Evaluation and statistical
  inference for human connectomes},}\ }\href@noop {} {\bibfield  {journal}
  {\bibinfo  {journal} {Nature methods}\ }\textbf {\bibinfo {volume} {11}},\
  \bibinfo {pages} {1058--1063} (\bibinfo {year} {2014})}\BibitemShut {NoStop}%
\bibitem [{\citenamefont {Oh}\ \emph {et~al.}(2014)\citenamefont {Oh},
  \citenamefont {Harris}, \citenamefont {Ng}, \citenamefont {Winslow},
  \citenamefont {Cain}, \citenamefont {Mihalas}, \citenamefont {Wang},
  \citenamefont {Lau}, \citenamefont {Kuan}, \citenamefont {Henry} \emph
  {et~al.}}]{oh2014mesoscale}%
  \BibitemOpen
  \bibfield  {author} {\bibinfo {author} {\bibfnamefont {Seung~Wook}\
  \bibnamefont {Oh}}, \bibinfo {author} {\bibfnamefont {Julie~A}\ \bibnamefont
  {Harris}}, \bibinfo {author} {\bibfnamefont {Lydia}\ \bibnamefont {Ng}},
  \bibinfo {author} {\bibfnamefont {Brent}\ \bibnamefont {Winslow}}, \bibinfo
  {author} {\bibfnamefont {Nicholas}\ \bibnamefont {Cain}}, \bibinfo {author}
  {\bibfnamefont {Stefan}\ \bibnamefont {Mihalas}}, \bibinfo {author}
  {\bibfnamefont {Quanxin}\ \bibnamefont {Wang}}, \bibinfo {author}
  {\bibfnamefont {Chris}\ \bibnamefont {Lau}}, \bibinfo {author} {\bibfnamefont
  {Leonard}\ \bibnamefont {Kuan}}, \bibinfo {author} {\bibfnamefont {Alex~M}\
  \bibnamefont {Henry}},  \emph {et~al.},\ }\bibfield  {title} {\enquote
  {\bibinfo {title} {A mesoscale connectome of the mouse brain},}\ }\href@noop
  {} {\bibfield  {journal} {\bibinfo  {journal} {Nature}\ }\textbf {\bibinfo
  {volume} {508}},\ \bibinfo {pages} {207} (\bibinfo {year}
  {2014})}\BibitemShut {NoStop}%
\bibitem [{\citenamefont {Chiang}\ \emph {et~al.}(2011)\citenamefont {Chiang},
  \citenamefont {Lin}, \citenamefont {Chuang}, \citenamefont {Chang},
  \citenamefont {Hsieh}, \citenamefont {Yeh}, \citenamefont {Shih},
  \citenamefont {Wu}, \citenamefont {Wang}, \citenamefont {Chen} \emph
  {et~al.}}]{chiang2011three}%
  \BibitemOpen
  \bibfield  {author} {\bibinfo {author} {\bibfnamefont {Ann-Shyn}\
  \bibnamefont {Chiang}}, \bibinfo {author} {\bibfnamefont {Chih-Yung}\
  \bibnamefont {Lin}}, \bibinfo {author} {\bibfnamefont {Chao-Chun}\
  \bibnamefont {Chuang}}, \bibinfo {author} {\bibfnamefont {Hsiu-Ming}\
  \bibnamefont {Chang}}, \bibinfo {author} {\bibfnamefont {Chang-Huain}\
  \bibnamefont {Hsieh}}, \bibinfo {author} {\bibfnamefont {Chang-Wei}\
  \bibnamefont {Yeh}}, \bibinfo {author} {\bibfnamefont {Chi-Tin}\ \bibnamefont
  {Shih}}, \bibinfo {author} {\bibfnamefont {Jian-Jheng}\ \bibnamefont {Wu}},
  \bibinfo {author} {\bibfnamefont {Guo-Tzau}\ \bibnamefont {Wang}}, \bibinfo
  {author} {\bibfnamefont {Yung-Chang}\ \bibnamefont {Chen}},  \emph {et~al.},\
  }\bibfield  {title} {\enquote {\bibinfo {title} {Three-dimensional
  reconstruction of brain-wide wiring networks in drosophila at single-cell
  resolution},}\ }\href@noop {} {\bibfield  {journal} {\bibinfo  {journal}
  {Current Biology}\ }\textbf {\bibinfo {volume} {21}},\ \bibinfo {pages}
  {1--11} (\bibinfo {year} {2011})}\BibitemShut {NoStop}%
\bibitem [{\citenamefont {Worrell}\ \emph {et~al.}(2017)\citenamefont
  {Worrell}, \citenamefont {Rumschlag}, \citenamefont {Betzel}, \citenamefont
  {Sporns},\ and\ \citenamefont {Mi{\v{s}}i{\'c}}}]{worrell2017optimized}%
  \BibitemOpen
  \bibfield  {author} {\bibinfo {author} {\bibfnamefont {Jacob~C}\ \bibnamefont
  {Worrell}}, \bibinfo {author} {\bibfnamefont {Jeffrey}\ \bibnamefont
  {Rumschlag}}, \bibinfo {author} {\bibfnamefont {Richard~F}\ \bibnamefont
  {Betzel}}, \bibinfo {author} {\bibfnamefont {Olaf}\ \bibnamefont {Sporns}}, \
  and\ \bibinfo {author} {\bibfnamefont {Bratislav}\ \bibnamefont
  {Mi{\v{s}}i{\'c}}},\ }\bibfield  {title} {\enquote {\bibinfo {title}
  {Optimized connectome architecture for sensorymotor integration},}\
  }\href@noop {} {\bibfield  {journal} {\bibinfo  {journal} {Network
  Neuroscience}\ } (\bibinfo {year} {2017})}\BibitemShut {NoStop}%
\bibitem [{\citenamefont {Yeh}\ \emph {et~al.}(2011)\citenamefont {Yeh},
  \citenamefont {Wedeen},\ and\ \citenamefont {Tseng}}]{yeh2011estimation}%
  \BibitemOpen
  \bibfield  {author} {\bibinfo {author} {\bibfnamefont {Fang-Cheng}\
  \bibnamefont {Yeh}}, \bibinfo {author} {\bibfnamefont {Van~Jay}\ \bibnamefont
  {Wedeen}}, \ and\ \bibinfo {author} {\bibfnamefont {Wen-Yih~Isaac}\
  \bibnamefont {Tseng}},\ }\bibfield  {title} {\enquote {\bibinfo {title}
  {Estimation of fiber orientation and spin density distribution by diffusion
  deconvolution},}\ }\href@noop {} {\bibfield  {journal} {\bibinfo  {journal}
  {Neuroimage}\ }\textbf {\bibinfo {volume} {55}},\ \bibinfo {pages}
  {1054--1062} (\bibinfo {year} {2011})}\BibitemShut {NoStop}%
\bibitem [{\citenamefont {Roberts}\ \emph {et~al.}(2017)\citenamefont
  {Roberts}, \citenamefont {Perry}, \citenamefont {Roberts}, \citenamefont
  {Mitchell},\ and\ \citenamefont {Breakspear}}]{roberts2017consistency}%
  \BibitemOpen
  \bibfield  {author} {\bibinfo {author} {\bibfnamefont {James~A}\ \bibnamefont
  {Roberts}}, \bibinfo {author} {\bibfnamefont {Alistair}\ \bibnamefont
  {Perry}}, \bibinfo {author} {\bibfnamefont {Gloria}\ \bibnamefont {Roberts}},
  \bibinfo {author} {\bibfnamefont {Philip~B}\ \bibnamefont {Mitchell}}, \ and\
  \bibinfo {author} {\bibfnamefont {Michael}\ \bibnamefont {Breakspear}},\
  }\bibfield  {title} {\enquote {\bibinfo {title} {Consistency-based
  thresholding of the human connectome},}\ }\href@noop {} {\bibfield  {journal}
  {\bibinfo  {journal} {Neuroimage}\ }\textbf {\bibinfo {volume} {145}},\
  \bibinfo {pages} {118--129} (\bibinfo {year} {2017})}\BibitemShut {NoStop}%
\bibitem [{\citenamefont {Dijkstra}(1959)}]{dijkstra1959note}%
  \BibitemOpen
  \bibfield  {author} {\bibinfo {author} {\bibfnamefont {Edsger~W}\
  \bibnamefont {Dijkstra}},\ }\bibfield  {title} {\enquote {\bibinfo {title} {A
  note on two problems in connexion with graphs},}\ }\href@noop {} {\bibfield
  {journal} {\bibinfo  {journal} {Numerische mathematik}\ }\textbf {\bibinfo
  {volume} {1}},\ \bibinfo {pages} {269--271} (\bibinfo {year}
  {1959})}\BibitemShut {NoStop}%
\bibitem [{\citenamefont {Brandes}(2001)}]{brandes2001faster}%
  \BibitemOpen
  \bibfield  {author} {\bibinfo {author} {\bibfnamefont {Ulrik}\ \bibnamefont
  {Brandes}},\ }\bibfield  {title} {\enquote {\bibinfo {title} {A faster
  algorithm for betweenness centrality},}\ }\href@noop {} {\bibfield  {journal}
  {\bibinfo  {journal} {Journal of mathematical sociology}\ }\textbf {\bibinfo
  {volume} {25}},\ \bibinfo {pages} {163--177} (\bibinfo {year}
  {2001})}\BibitemShut {NoStop}%
\bibitem [{\citenamefont {Hilgetag}\ \emph {et~al.}(2002)\citenamefont
  {Hilgetag}, \citenamefont {K{\"o}tter}, \citenamefont {Stephan},\ and\
  \citenamefont {Sporns}}]{hilgetag2002computational}%
  \BibitemOpen
  \bibfield  {author} {\bibinfo {author} {\bibfnamefont {Claus~C}\ \bibnamefont
  {Hilgetag}}, \bibinfo {author} {\bibfnamefont {Rolf}\ \bibnamefont
  {K{\"o}tter}}, \bibinfo {author} {\bibfnamefont {Klaas~E}\ \bibnamefont
  {Stephan}}, \ and\ \bibinfo {author} {\bibfnamefont {Olaf}\ \bibnamefont
  {Sporns}},\ }\bibfield  {title} {\enquote {\bibinfo {title} {Computational
  methods for the analysis of brain connectivity},}\ }in\ \href@noop {} {\emph
  {\bibinfo {booktitle} {Computational neuroanatomy}}}\ (\bibinfo  {publisher}
  {Springer},\ \bibinfo {year} {2002})\ pp.\ \bibinfo {pages}
  {295--335}\BibitemShut {NoStop}%
\bibitem [{\citenamefont {Barbas}\ and\ \citenamefont
  {Pandya}(1989)}]{barbas1989architecture}%
  \BibitemOpen
  \bibfield  {author} {\bibinfo {author} {\bibfnamefont {H}~\bibnamefont
  {Barbas}}\ and\ \bibinfo {author} {\bibfnamefont {DN}~\bibnamefont
  {Pandya}},\ }\bibfield  {title} {\enquote {\bibinfo {title} {Architecture and
  intrinsic connections of the prefrontal cortex in the rhesus monkey},}\
  }\href@noop {} {\bibfield  {journal} {\bibinfo  {journal} {Journal of
  Comparative Neurology}\ }\textbf {\bibinfo {volume} {286}},\ \bibinfo {pages}
  {353--375} (\bibinfo {year} {1989})}\BibitemShut {NoStop}%
\bibitem [{\citenamefont {Young}(1993)}]{young1993organization}%
  \BibitemOpen
  \bibfield  {author} {\bibinfo {author} {\bibfnamefont {Malcolm~P}\
  \bibnamefont {Young}},\ }\bibfield  {title} {\enquote {\bibinfo {title} {The
  organization of neural systems in the primate cerebral cortex},}\ }\href@noop
  {} {\bibfield  {journal} {\bibinfo  {journal} {Proceedings of the Royal
  Society of London B: Biological Sciences}\ }\textbf {\bibinfo {volume}
  {252}},\ \bibinfo {pages} {13--18} (\bibinfo {year} {1993})}\BibitemShut
  {NoStop}%
\bibitem [{\citenamefont {Scannell}\ \emph {et~al.}(1995)\citenamefont
  {Scannell}, \citenamefont {Blakemore},\ and\ \citenamefont
  {Young}}]{scannell1995analysis}%
  \BibitemOpen
  \bibfield  {author} {\bibinfo {author} {\bibfnamefont {Jack~W}\ \bibnamefont
  {Scannell}}, \bibinfo {author} {\bibfnamefont {Colin}\ \bibnamefont
  {Blakemore}}, \ and\ \bibinfo {author} {\bibfnamefont {Malcolm~P}\
  \bibnamefont {Young}},\ }\bibfield  {title} {\enquote {\bibinfo {title}
  {Analysis of connectivity in the cat cerebral cortex},}\ }\href@noop {}
  {\bibfield  {journal} {\bibinfo  {journal} {Journal of Neuroscience}\
  }\textbf {\bibinfo {volume} {15}},\ \bibinfo {pages} {1463--1483} (\bibinfo
  {year} {1995})}\BibitemShut {NoStop}%
\bibitem [{\citenamefont {Albert}\ \emph {et~al.}(2000)\citenamefont {Albert},
  \citenamefont {Jeong},\ and\ \citenamefont {Barab{\'a}si}}]{albert2000error}%
  \BibitemOpen
  \bibfield  {author} {\bibinfo {author} {\bibfnamefont {R{\'e}ka}\
  \bibnamefont {Albert}}, \bibinfo {author} {\bibfnamefont {Hawoong}\
  \bibnamefont {Jeong}}, \ and\ \bibinfo {author} {\bibfnamefont
  {Albert-L{\'a}szl{\'o}}\ \bibnamefont {Barab{\'a}si}},\ }\bibfield  {title}
  {\enquote {\bibinfo {title} {Error and attack tolerance of complex
  networks},}\ }\href@noop {} {\bibfield  {journal} {\bibinfo  {journal} {arXiv
  preprint cond-mat/0008064}\ } (\bibinfo {year} {2000})}\BibitemShut {NoStop}%
\bibitem [{\citenamefont {Tononi}\ \emph {et~al.}(1999)\citenamefont {Tononi},
  \citenamefont {Sporns},\ and\ \citenamefont {Edelman}}]{tononi1999measures}%
  \BibitemOpen
  \bibfield  {author} {\bibinfo {author} {\bibfnamefont {Giulio}\ \bibnamefont
  {Tononi}}, \bibinfo {author} {\bibfnamefont {Olaf}\ \bibnamefont {Sporns}}, \
  and\ \bibinfo {author} {\bibfnamefont {Gerald~M}\ \bibnamefont {Edelman}},\
  }\bibfield  {title} {\enquote {\bibinfo {title} {Measures of degeneracy and
  redundancy in biological networks},}\ }\href@noop {} {\bibfield  {journal}
  {\bibinfo  {journal} {Proceedings of the National Academy of Sciences}\
  }\textbf {\bibinfo {volume} {96}},\ \bibinfo {pages} {3257--3262} (\bibinfo
  {year} {1999})}\BibitemShut {NoStop}%
\bibitem [{\citenamefont {Newman}(2012)}]{newman2012communities}%
  \BibitemOpen
  \bibfield  {author} {\bibinfo {author} {\bibfnamefont {Mark~EJ}\ \bibnamefont
  {Newman}},\ }\bibfield  {title} {\enquote {\bibinfo {title} {Communities,
  modules and large-scale structure in networks},}\ }\href@noop {} {\bibfield
  {journal} {\bibinfo  {journal} {Nature physics}\ }\textbf {\bibinfo {volume}
  {8}},\ \bibinfo {pages} {25} (\bibinfo {year} {2012})}\BibitemShut {NoStop}%
\bibitem [{\citenamefont {Sporns}\ and\ \citenamefont
  {Betzel}(2016)}]{sporns2016modular}%
  \BibitemOpen
  \bibfield  {author} {\bibinfo {author} {\bibfnamefont {Olaf}\ \bibnamefont
  {Sporns}}\ and\ \bibinfo {author} {\bibfnamefont {Richard~F}\ \bibnamefont
  {Betzel}},\ }\bibfield  {title} {\enquote {\bibinfo {title} {Modular brain
  networks},}\ }\href@noop {} {\bibfield  {journal} {\bibinfo  {journal}
  {Annual review of psychology}\ }\textbf {\bibinfo {volume} {67}},\ \bibinfo
  {pages} {613--640} (\bibinfo {year} {2016})}\BibitemShut {NoStop}%
\bibitem [{\citenamefont {Newman}\ and\ \citenamefont
  {Girvan}(2004)}]{newman2004finding}%
  \BibitemOpen
  \bibfield  {author} {\bibinfo {author} {\bibfnamefont {Mark~EJ}\ \bibnamefont
  {Newman}}\ and\ \bibinfo {author} {\bibfnamefont {Michelle}\ \bibnamefont
  {Girvan}},\ }\bibfield  {title} {\enquote {\bibinfo {title} {Finding and
  evaluating community structure in networks},}\ }\href@noop {} {\bibfield
  {journal} {\bibinfo  {journal} {Physical review E}\ }\textbf {\bibinfo
  {volume} {69}},\ \bibinfo {pages} {026113} (\bibinfo {year}
  {2004})}\BibitemShut {NoStop}%
\bibitem [{\citenamefont {Leicht}\ and\ \citenamefont
  {Newman}(2008)}]{leicht2008community}%
  \BibitemOpen
  \bibfield  {author} {\bibinfo {author} {\bibfnamefont {Elizabeth~A}\
  \bibnamefont {Leicht}}\ and\ \bibinfo {author} {\bibfnamefont {Mark~EJ}\
  \bibnamefont {Newman}},\ }\bibfield  {title} {\enquote {\bibinfo {title}
  {Community structure in directed networks},}\ }\href@noop {} {\bibfield
  {journal} {\bibinfo  {journal} {Physical review letters}\ }\textbf {\bibinfo
  {volume} {100}},\ \bibinfo {pages} {118703} (\bibinfo {year}
  {2008})}\BibitemShut {NoStop}%
\bibitem [{\citenamefont {Jutla}\ \emph {et~al.}(2011)\citenamefont {Jutla},
  \citenamefont {Jeub},\ and\ \citenamefont {Mucha}}]{jutla2011generalized}%
  \BibitemOpen
  \bibfield  {author} {\bibinfo {author} {\bibfnamefont {Inderjit~S}\
  \bibnamefont {Jutla}}, \bibinfo {author} {\bibfnamefont {Lucas~GS}\
  \bibnamefont {Jeub}}, \ and\ \bibinfo {author} {\bibfnamefont {Peter~J}\
  \bibnamefont {Mucha}},\ }\bibfield  {title} {\enquote {\bibinfo {title} {A
  generalized louvain method for community detection implemented in matlab},}\
  }\href@noop {} {\bibfield  {journal} {\bibinfo  {journal} {URL
  http://netwiki. amath. unc. edu/GenLouvain}\ } (\bibinfo {year}
  {2011})}\BibitemShut {NoStop}%
\bibitem [{\citenamefont {Traud}\ \emph {et~al.}(2011)\citenamefont {Traud},
  \citenamefont {Kelsic}, \citenamefont {Mucha},\ and\ \citenamefont
  {Porter}}]{traud2011comparing}%
  \BibitemOpen
  \bibfield  {author} {\bibinfo {author} {\bibfnamefont {Amanda~L}\
  \bibnamefont {Traud}}, \bibinfo {author} {\bibfnamefont {Eric~D}\
  \bibnamefont {Kelsic}}, \bibinfo {author} {\bibfnamefont {Peter~J}\
  \bibnamefont {Mucha}}, \ and\ \bibinfo {author} {\bibfnamefont {Mason~A}\
  \bibnamefont {Porter}},\ }\bibfield  {title} {\enquote {\bibinfo {title}
  {Comparing community structure to characteristics in online collegiate social
  networks},}\ }\href@noop {} {\bibfield  {journal} {\bibinfo  {journal} {SIAM
  review}\ }\textbf {\bibinfo {volume} {53}},\ \bibinfo {pages} {526--543}
  (\bibinfo {year} {2011})}\BibitemShut {NoStop}%
\bibitem [{\citenamefont {Betzel}\ \emph {et~al.}(2013)\citenamefont {Betzel},
  \citenamefont {Griffa}, \citenamefont {Avena-Koenigsberger}, \citenamefont
  {Go{\~n}i}, \citenamefont {Thiran}, \citenamefont {Hagmann},\ and\
  \citenamefont {Sporns}}]{betzel2013multi}%
  \BibitemOpen
  \bibfield  {author} {\bibinfo {author} {\bibfnamefont {Richard~F}\
  \bibnamefont {Betzel}}, \bibinfo {author} {\bibfnamefont {Alessandra}\
  \bibnamefont {Griffa}}, \bibinfo {author} {\bibfnamefont {Andrea}\
  \bibnamefont {Avena-Koenigsberger}}, \bibinfo {author} {\bibfnamefont
  {Joaqu{\'\i}n}\ \bibnamefont {Go{\~n}i}}, \bibinfo {author} {\bibfnamefont
  {Jean-Philippe}\ \bibnamefont {Thiran}}, \bibinfo {author} {\bibfnamefont
  {Patric}\ \bibnamefont {Hagmann}}, \ and\ \bibinfo {author} {\bibfnamefont
  {Olaf}\ \bibnamefont {Sporns}},\ }\bibfield  {title} {\enquote {\bibinfo
  {title} {Multi-scale community organization of the human structural
  connectome and its relationship with resting-state functional
  connectivity},}\ }\href@noop {} {\bibfield  {journal} {\bibinfo  {journal}
  {Network Science}\ }\textbf {\bibinfo {volume} {1}},\ \bibinfo {pages}
  {353--373} (\bibinfo {year} {2013})}\BibitemShut {NoStop}%
\bibitem [{\citenamefont {Mucha}\ \emph {et~al.}(2010)\citenamefont {Mucha},
  \citenamefont {Richardson}, \citenamefont {Macon}, \citenamefont {Porter},\
  and\ \citenamefont {Onnela}}]{mucha2010community}%
  \BibitemOpen
  \bibfield  {author} {\bibinfo {author} {\bibfnamefont {Peter~J}\ \bibnamefont
  {Mucha}}, \bibinfo {author} {\bibfnamefont {Thomas}\ \bibnamefont
  {Richardson}}, \bibinfo {author} {\bibfnamefont {Kevin}\ \bibnamefont
  {Macon}}, \bibinfo {author} {\bibfnamefont {Mason~A}\ \bibnamefont {Porter}},
  \ and\ \bibinfo {author} {\bibfnamefont {Jukka-Pekka}\ \bibnamefont
  {Onnela}},\ }\bibfield  {title} {\enquote {\bibinfo {title} {Community
  structure in time-dependent, multiscale, and multiplex networks},}\
  }\href@noop {} {\bibfield  {journal} {\bibinfo  {journal} {science}\ }\textbf
  {\bibinfo {volume} {328}},\ \bibinfo {pages} {876--878} (\bibinfo {year}
  {2010})}\BibitemShut {NoStop}%
\bibitem [{\citenamefont {MacMahon}\ and\ \citenamefont
  {Garlaschelli}(2013)}]{macmahon2013community}%
  \BibitemOpen
  \bibfield  {author} {\bibinfo {author} {\bibfnamefont {Mel}\ \bibnamefont
  {MacMahon}}\ and\ \bibinfo {author} {\bibfnamefont {Diego}\ \bibnamefont
  {Garlaschelli}},\ }\bibfield  {title} {\enquote {\bibinfo {title} {Community
  detection for correlation matrices},}\ }\href@noop {} {\bibfield  {journal}
  {\bibinfo  {journal} {arXiv preprint arXiv:1311.1924}\ } (\bibinfo {year}
  {2013})}\BibitemShut {NoStop}%
\bibitem [{\citenamefont {Bazzi}\ \emph {et~al.}(2016)\citenamefont {Bazzi},
  \citenamefont {Porter}, \citenamefont {Williams}, \citenamefont {McDonald},
  \citenamefont {Fenn},\ and\ \citenamefont {Howison}}]{bazzi2016community}%
  \BibitemOpen
  \bibfield  {author} {\bibinfo {author} {\bibfnamefont {Marya}\ \bibnamefont
  {Bazzi}}, \bibinfo {author} {\bibfnamefont {Mason~A}\ \bibnamefont {Porter}},
  \bibinfo {author} {\bibfnamefont {Stacy}\ \bibnamefont {Williams}}, \bibinfo
  {author} {\bibfnamefont {Mark}\ \bibnamefont {McDonald}}, \bibinfo {author}
  {\bibfnamefont {Daniel~J}\ \bibnamefont {Fenn}}, \ and\ \bibinfo {author}
  {\bibfnamefont {Sam~D}\ \bibnamefont {Howison}},\ }\bibfield  {title}
  {\enquote {\bibinfo {title} {Community detection in temporal multilayer
  networks, with an application to correlation networks},}\ }\href@noop {}
  {\bibfield  {journal} {\bibinfo  {journal} {Multiscale Modeling \&
  Simulation}\ }\textbf {\bibinfo {volume} {14}},\ \bibinfo {pages} {1--41}
  (\bibinfo {year} {2016})}\BibitemShut {NoStop}%
\bibitem [{\citenamefont {Guimera}\ and\ \citenamefont
  {Amaral}(2005)}]{guimera2005functional}%
  \BibitemOpen
  \bibfield  {author} {\bibinfo {author} {\bibfnamefont {Roger}\ \bibnamefont
  {Guimera}}\ and\ \bibinfo {author} {\bibfnamefont {Luis A~Nunes}\
  \bibnamefont {Amaral}},\ }\bibfield  {title} {\enquote {\bibinfo {title}
  {Functional cartography of complex metabolic networks},}\ }\href@noop {}
  {\bibfield  {journal} {\bibinfo  {journal} {nature}\ }\textbf {\bibinfo
  {volume} {433}},\ \bibinfo {pages} {895} (\bibinfo {year}
  {2005})}\BibitemShut {NoStop}%
\bibitem [{\citenamefont {Bertolero}\ \emph {et~al.}(2017)\citenamefont
  {Bertolero}, \citenamefont {Yeo},\ and\ \citenamefont
  {D'Esposito}}]{bertolero2017diverse}%
  \BibitemOpen
  \bibfield  {author} {\bibinfo {author} {\bibfnamefont {MA}~\bibnamefont
  {Bertolero}}, \bibinfo {author} {\bibfnamefont {BTT}\ \bibnamefont {Yeo}}, \
  and\ \bibinfo {author} {\bibfnamefont {M}~\bibnamefont {D'Esposito}},\
  }\bibfield  {title} {\enquote {\bibinfo {title} {The diverse club: The
  integrative core of complex networks},}\ }\href@noop {} {\bibfield  {journal}
  {\bibinfo  {journal} {arXiv preprint arXiv:1701.01150}\ } (\bibinfo {year}
  {2017})}\BibitemShut {NoStop}%
\bibitem [{\citenamefont {Woolrich}\ and\ \citenamefont
  {Stephan}(2013)}]{woolrich2013biophysical}%
  \BibitemOpen
  \bibfield  {author} {\bibinfo {author} {\bibfnamefont {Mark~W}\ \bibnamefont
  {Woolrich}}\ and\ \bibinfo {author} {\bibfnamefont {Klaas~E}\ \bibnamefont
  {Stephan}},\ }\bibfield  {title} {\enquote {\bibinfo {title} {Biophysical
  network models and the human connectome},}\ }\href@noop {} {\bibfield
  {journal} {\bibinfo  {journal} {Neuroimage}\ }\textbf {\bibinfo {volume}
  {80}},\ \bibinfo {pages} {330--338} (\bibinfo {year} {2013})}\BibitemShut
  {NoStop}%
\end{thebibliography}%
	
\end{document}